# Simulation of snakes traversing a wedge obstacle using vertical body bending

Yifeng Zhang, Qihan Xuan, Qiyuan Fu, *Chen Li

Department of Mechanical Engineering, Johns Hopkins University

*Corresponding author. Email: [redacted]

## Abstract

Snakes are excellent at slithering across a variety of terrain. Many studies focused on how snakes use lateral body bending to generate propulsion against asperities on flat surfaces. Only recently did we realize that snakes can also use vertical bending to propel against terrain of varying height, such as a horizontal ladder and a wedge (Jurestovsky, Usher, Astley, 2021, JEB) or uneven terrain (Fu, Astley, Li, 2021, SICB). Here, to understand how to use vertical bending to generate propulsion, we developed a dynamic simulation of snakes traversing a wedge obstacle (slope = 27°, length ≈ 0.1 body length, similar to animal experiment), using a recent numerical discrete elastic rods method (Zhang et al, 2019, *Nat. Comm.*). The snake was modeled as an elastic Cosserat rod, which can bend itself by internal forces. The interaction between the snake and terrain was modeled via a spring-damper model for normal force plus Coulomb friction. The rod is governed by internal forces including tensile/compressive, shear, bending, twisting forces due to deformation, internal torque given by a controller, and external forces including terrain reaction forces and gravitational force. We found that a posteriorly propagating internal torque profile with a maximum on body segments around the wedge obstacle generated near steady-speed forward locomotion as observed in the animal, with torque magnitude insensitive to locomotion speed. Remarkably, for a snake-terrain kinetic friction coefficient of only 0.20, the body had to push into the sloped wedge surface with a pressure as high as 5 times that from body weight to generate sufficient forward propulsion to overcome frictional drag. This suggested that snakes have a large capacity to use vertical bending to push against the environment to generate propulsion. We are performing systematic parameter variation in our simulation to discover propulsion principles.

## Introduction

Snakes are excellent at slithering across a variety of terrain. Unlike limbed vertebrates which normally have discrete contact points with the ground, snakes remain continuous contact with the ground to provide propulsion and maintain stability. The frictional force from the terrain that overwhelms the inertial effects renders the movement. There are many kinds of motion patterns found in biological snakes, including lateral undulation, concertina, sidewinding, and rectilinear crawling. Among these, lateral undulation is one of the most common and energy-efficient locomotion modes for snakes. It is a continuous movement of the entire body propagating waves from the anterior to the posterior body. Findings indicate that their skin holds the anisotropic force against the terrain (Hu et al., 2009) , which is crucial to this locomotion mode while on flat ground. Without utilizing the anisotropic properties, snakes may use contours such as rocks in the environment to push against (Kano and Ishiguro, 2013; Schiebel et al., 2020; Travers et al., 2018). It is found that snake can use lateral body bending to actively push against vertical structures to generate propulsion against asperities on flat surfaces and was tested by robophysical model. However, only recently did researchers realize that snakes can also use vertical bending to propel against terrain of varying heights, such as a horizontal ladder and a wedge (Jurestovsky et al., 2021) or uneven terrain (Fu et al., 2022) via a similar mechanism used in lateral bending. The snake looks simply propagating its body shape backward. The force pattern measured in the experiment (Jurestovsky et al., 2021) shows that the snake can generate fore-aft propulsion against the environment to achieve a forward motion. Robots (Jurestovsky et al., 2021) were fabricated to verify that snakes can use only vertical bending to provide enough propulsion. There is also one snake robot (Date and Takita, 2005) using vertical bending to traverse a single cylindrical obstacle with an oversimplified model for control.

However, it is difficult to measure forces in animal experiments. It is because that the external forces require force sensor with high resolution and wide spatial distribution, and the internal force of the animal is impossible to measure by conventional force sensor. Proper computational modeling for animal dynamics can help predict the internal forces and interaction forces with the terrain, as well as elucidate the locomotion mode (Jayne, 2020). Modeling the dynamics of the snake locomotion helps researchers better understand the snake's motion. A number of modeling work has been proposed, including rigid body modeling and continuous modeling (Transeth et al., 2009). Due to the similarity between the snake-like robot and the snake, massive modeling work from the robot can be a good reference for animal modeling. The snake robot dynamics are commonly analyzed by the muti-link model. Other than traditional kinematics based on D-H convention, backbone curves (Fu et al., 2021) are also employed to reflect the features of the hyper-redundant robot, which inversely inspired the snake kinematics. Not like traditional wheeled snake robots, whose dynamics can be solved by kinematics constraints, the interaction between the robots and the terrain plays an important role. Friction models presented in the literature are primarily based on a Coulomb or viscous-like friction model. Also, it is necessary to model the normal contact force due to impacts and sustained contact with the terrain. It has been tested that this force can be characterized as compliant by a spring-damper model (Liljebäck et al., 2008).

Similar to modeling of snake robot, a biological snake can be modeled by numerous short links. The friction between the snake significantly affects its motion. The skin surface of the snake plays an important role in the magnitude of the friction, which gives the snake higher transversal and backward friction. A model proposed by (Hu et al., 2009), which described friction anisotropy, is proved to be effective. However, little did previous research use a computational model to analyze the vertical bending of snakes.

In our work, to understand how to use vertical bending to generate propulsion, we developed a dynamic simulation of snakes traversing a wedge obstacle similar to the animal experiment, using a recent numerical discrete elastic rods method (Gazzola et al., 2018). The Cosserat rod enables all six degrees of freedom and contains elastic forces. The key assumption when modeling rods are that their length is much larger than their radius. Hence, the snake can be modeled as a single Cosserat rod for simplicity. Inspired by the animal experiment observation that the snake propagates its body shape backward, we set this kind of pure propagation in our computational model to exhibit the snake's motion. Next, with the help of computational mechanics, we can run the simulation of a snake traversing obstacles of different sizes and terrain and output the force patterns, which are all hard to measure in animal experiments. We ran simulations hundreds of times to characterize the major factors that affect the vertical bending locomotion. Features of the terrain and snake are tuned extensively, including friction coefficient, height and the size of the obstacle, etc. We placed the snake in different locations to test the effect of the initial location of the snake relative to the obstacle which proved to be important. Finally, we explored different ways of locomotion control. In order to achieve a steady motion, force feedback control is needed to stabilize the motion. We compare results with different control strategies applied.

# Methods

## Mathematical description of a single Cosserat rod

We used a single Cosserat rod to model the slender body of the snake. The Cosserat rod is governed by internal forces, including extension/compression, shearing, bending, and twisting forces due to deformation.

We recall the mathematical basis and numerical methods of the modeling of the Cosserat rod. The filament can be represented by a centerline $r$ and an oriented frame of reference $\boldsymbol{Q}$. The angular velocity and curvature in the body-convected frame can be represented in terms of $\boldsymbol{Q}$, respectively

$$\boldsymbol{\omega}_{\mathcal{L}} = \left( \boldsymbol{Q} \frac{\partial \boldsymbol{Q}^T}{\partial t} \right)^{\vee}$$

$$\boldsymbol{\kappa}_{\mathcal{L}} = \left(\boldsymbol{Q}\frac{\partial \boldsymbol{Q}^T}{\partial s}\right)^{\vee}$$

Here, the disjunction operator denotes the conversion from the skew-symmetric matrix to its associated vector. The dynamics of a slender elastic body are described in Eulerian-Lagrangian form:

$$\begin{cases} \dfrac{\partial \boldsymbol{r}}{\partial t} = \boldsymbol{v} \\ \dfrac{\partial \boldsymbol{d}_j}{\partial t} = (\boldsymbol{Q}^T \boldsymbol{\omega}_{\mathcal{L}}) \times \boldsymbol{d}_j, j = 1,2,3 \\ \dfrac{\partial (\rho A \boldsymbol{v})}{\partial t} = \dfrac{\partial (\boldsymbol{Q}^T \boldsymbol{n}_{\mathcal{L}})}{\partial s} + \boldsymbol{f} \\ \dfrac{\partial (\rho \boldsymbol{I} \boldsymbol{\omega}_{\mathcal{L}})}{\partial t} = \dfrac{\partial \boldsymbol{\tau}_{\mathcal{L}}}{\partial s} + \boldsymbol{\kappa}_{\mathcal{L}} \times \boldsymbol{\tau}_{\mathcal{L}} + \boldsymbol{Q}\dfrac{\partial \boldsymbol{r}}{\partial s} \times \boldsymbol{n}_{\mathcal{L}} + (\rho \boldsymbol{I} \boldsymbol{\omega}_{\mathcal{L}}) \times \boldsymbol{\omega}_{\mathcal{L}} + \boldsymbol{c}_{\mathcal{L}} \end{cases}$$

Where $\boldsymbol{r}$ is the position, $\boldsymbol{v}$ is the velocity, $\boldsymbol{n}_{\mathcal{L}}$ is the internal force, $\boldsymbol{\tau}_{\mathcal{L}}$ is the internal torque. $\boldsymbol{f}$ is the external body force line density, and $\boldsymbol{c}$ is the external body torque line density. The tensor $\boldsymbol{I}$ is the second area moment of inertia (We assume circular cross-sections). Note that the second area moment of inertia is subject to the limit condition of the element length approaching zero. The strains can be expressed by the vector $\boldsymbol{\sigma}_{\mathcal{L}} = \boldsymbol{Q}\left(\frac{\partial r}{\partial \hat{s}} - \boldsymbol{d}_3\right) = \boldsymbol{Q}(e\boldsymbol{t} - \boldsymbol{d}_3)$, where the unit tangential vector $\boldsymbol{t} = \frac{\partial \boldsymbol{r}}{\partial s} = \frac{1}{e} \cdot \frac{\partial \boldsymbol{r}}{\partial \hat{s}}$. Here, $\hat{s}$ denotes the unstretched coordinate of the slender body. The scalar field measuring the axial strain is hereby given by $e(\hat{s}, t) = ds/d\hat{s}$. Due to the axial strain, other parameters are updated based on the assumption that the material is incompressible:

$A = \frac{\hat{A}}{e}, \boldsymbol{I} = \frac{\hat{\boldsymbol{I}}}{e^2}, \boldsymbol{B} = \frac{\hat{\boldsymbol{B}}}{e^2}, \boldsymbol{S} = \frac{\hat{\boldsymbol{S}}}{e}$, and $\boldsymbol{\kappa}_{\mathcal{L}} = \frac{\widehat{\boldsymbol{\kappa}_{\mathcal{L}}}}{e}$

Where $A$ is the cross-section area, $\boldsymbol{B}$ is the bending rigidity, $\boldsymbol{S}$ is the shearing rigidity, $\boldsymbol{\kappa}_{\mathcal{L}}$ is the curvature. With that, the internal force and torque can be given by $\boldsymbol{n}_{\mathcal{L}} = \boldsymbol{S}\boldsymbol{\sigma}_{\mathcal{L}}$ and $\boldsymbol{\tau}_{\mathcal{L}} = \boldsymbol{B}\boldsymbol{\kappa}_{\mathcal{L}}$ assuming that the rod is not prestressed, respectively. Plugging that into the Eulerian-Lagrangian form, we can replace some variables with constants in terms of the axial strain $e$:

$$
\begin{cases}
\dfrac{\partial \boldsymbol{r}}{\partial t} = \boldsymbol{v} \\
\dfrac{\partial \boldsymbol{d}_j}{\partial t} = (\boldsymbol{Q}^T \boldsymbol{\omega}_{\mathcal{L}}) \times \boldsymbol{d}_j, j = 1,2,3 \\
\mathrm{d}m \cdot \dfrac{\partial \boldsymbol{v}}{\partial t} = \dfrac{\partial}{\partial \hat{s}} \left( \dfrac{\boldsymbol{Q}^T \widehat{\boldsymbol{S}} \boldsymbol{\sigma}_{\mathcal{L}}}{e} \right) \mathrm{d}\hat{s} + \mathrm{d}\boldsymbol{F} \\
\dfrac{\mathrm{d}\widehat{\boldsymbol{J}}}{e} \cdot \dfrac{\partial \boldsymbol{\omega}_{\mathcal{L}}}{\partial t} = \dfrac{\partial}{\partial \hat{s}} \left( \dfrac{\widehat{\boldsymbol{B}} \widehat{\boldsymbol{\kappa}}_{\mathcal{L}}}{e^3} \right) \mathrm{d}\hat{s} + \dfrac{\widehat{\boldsymbol{\kappa}}_{\mathcal{L}} \times \widehat{\boldsymbol{B}} \widehat{\boldsymbol{\kappa}}_{\mathcal{L}}}{e^3} \mathrm{d}\hat{s} + \left( \boldsymbol{Q} \boldsymbol{t} \times \widehat{\boldsymbol{S}} \boldsymbol{\sigma}_{\mathcal{L}} \right) \mathrm{d}\hat{s} + \left( \mathrm{d}\widehat{\boldsymbol{J}} \cdot \dfrac{\boldsymbol{\omega}_{\mathcal{L}}}{e} \right) \times \boldsymbol{\omega}_{\mathcal{L}} + \dfrac{\mathrm{d}\widehat{\boldsymbol{J}} \boldsymbol{\omega}_{\mathcal{L}}}{e^2} \cdot \dfrac{\partial e}{\partial t} + \mathrm{d}\boldsymbol{C}_{\mathcal{L}}
\end{cases}
$$

Here, $\mathrm{d}\boldsymbol{F}$ and $\mathrm{d}\boldsymbol{C}_{\mathcal{L}}$ represent the external force and external torque, respectively. The angular momentum equation composes the bend/twist internal couple, shear/stretch internal couple, Lagrangian transport, unsteady dilatation, and external couple. The next step is to discretize the spatial terms. The rod can be represented by a set of vertices $r_i(t)$, where $i = 1, \dots, N+1$. The centerline direction $\boldsymbol{t}_i$ can hereby be computed from the vertices by $\boldsymbol{t}_i = \frac{\boldsymbol{r}_{i+1} - \boldsymbol{r}_i}{\|\boldsymbol{r}_{i+1} - \boldsymbol{r}_i\|}$, where $i = 1, \dots, N$. In the modeling of the Cosserat rod, the centerline direction is not aligned with the normal direction of the cross-section. The difference between these two values leads to the shear force. To express the direction of the cross section, the set of rotation matrix of the material frames are denoted by $Q_i(t)$, where $i = 1, \dots, N$. In terms of the segments, the bending rigidity and the shearing rigidity can be discretized as $\widehat{\boldsymbol{B}}_{seg,i}$ and $\widehat{\boldsymbol{S}}_{seg,i}$ , where $i = 1, \dots, N$. As for the local stretching or compression ratio $e$, it can be discretized in terms of the segments

$$
e_i = \frac{\ell_i}{\hat{\ell}_i}, i = 1, \dots, N
$$

Where $\ell_i = \|\boldsymbol{r}_{i+1} - \boldsymbol{r}_i\|$ $(i = 1, \dots, N)$ is the length of the segment, $\hat{\ell}_i$ $(i = 1, \dots, N)$ is the original length of the segment. The local stretching or compression ratio $e$ can be discretized in terms of the elements

$$
\varepsilon_i = \frac{\ell_{i-1} + \ell_i}{2\hat{l}_i} = \frac{e_{i-1}\hat{\ell}_{i-1} + e_i \hat{\ell}_i}{2\hat{l}_i}, i = 2, \dots, N
$$

The first element and the last element are $\varepsilon_1 = \frac{\hat{\ell}_1}{2\hat{l}_1}$ and $\varepsilon_N = \frac{\hat{\ell}_N}{2\hat{l}_N}$. Where $\hat{l}_i$ $(i = 1, \dots, N+1)$ is the original length of the element. With similar manner, we can discretize the bending rigidity in terms of the elements

$$\widehat{\boldsymbol{B}}_{ele,i} = \frac{\widehat{\boldsymbol{B}}_{seg,i-1}\hat{\ell}_{i-1} + \widehat{\boldsymbol{B}}_{seg,i}\hat{\ell}_i}{2\hat{l}_i}, i = 1, \dots, N+1$$

The local curvature can be discretized as $\widehat{\boldsymbol{\kappa}}_{\mathcal{L}}^i = \frac{\log(\boldsymbol{Q}_{i-1}\boldsymbol{Q}_i^{\mathrm{T}})}{\hat{l}_i}$ in terms of the segments, where $i = 2, \dots, N$. The strain $\boldsymbol{\sigma}_{\mathcal{L},i} = e_i\boldsymbol{t}_i - \boldsymbol{d}_3$. The internal force term in equation xxx(c) can be discretized in terms of the segments

$$\boldsymbol{F}_{int,i} = \frac{\boldsymbol{Q}_i^T\widehat{\boldsymbol{S}}_{seg,i}\boldsymbol{\sigma}_{\mathcal{L},i}}{\varepsilon_i}, i = 1, \dots, N$$

Hence, equation xxx(c) evolves to

$$m_i \cdot \frac{\partial \boldsymbol{v}_i}{\partial t} = \Delta^{\mathrm{h}}\left(\boldsymbol{F}_{int,i}\right) + \mathrm{d}\boldsymbol{F}_i$$

where $\Delta^{\mathrm{h}}: \{\mathbb{R}^3\}_N \to \{\mathbb{R}^3\}_N$ is the discrete difference operator. Its definition is

$$y_{j=1,\dots,N+1} = \Delta^h\left(x_{i=1,\dots,N}\right) = \begin{cases} x_1, \text{ if } j = 1 \\ x_j - x_{j-1}, \text{ if } 1 < j \leq N \\ -x_N, \text{ if } j = N+1 \end{cases}$$

The equation xxx(d) which represents the angular momentum evolution can be discretized in a similar manner, albeit more difficult.

$$\frac{\hat{\boldsymbol{J}}_i}{e_i} \cdot \frac{\partial \boldsymbol{\omega}_{\mathcal{L}}^i}{\partial t} = \Delta^h\left(\frac{\widehat{\boldsymbol{B}}_i\widehat{\boldsymbol{\kappa}}_{\mathcal{L}}^i}{{\varepsilon_i}^3}\right) + \mathcal{A}^h\left(\frac{\widehat{\boldsymbol{\kappa}}_{\mathcal{L}}^i \times \widehat{\boldsymbol{B}}_{ele,i}\widehat{\boldsymbol{\kappa}}_{\mathcal{L}}^i}{{\varepsilon_i}^3} l_i\right) + \left(\boldsymbol{Q}_i\boldsymbol{t}_i \times \widehat{\boldsymbol{S}}_i\boldsymbol{\sigma}_{\mathcal{L}}^i\right)\hat{l}_i + \left(\hat{\boldsymbol{J}}_i \cdot \frac{\boldsymbol{\omega}_{\mathcal{L}}^i}{e_i}\right) \times \boldsymbol{\omega}_{\mathcal{L}}^i + \frac{\hat{\boldsymbol{J}}_i\boldsymbol{\omega}_{\mathcal{L}}^i}{e_i^2} \cdot \frac{\partial e_i}{\partial t} + \boldsymbol{C}_{\mathcal{L}}^i$$

where $\mathcal{A}^h: \{\mathbb{R}^3\}_N \to \{\mathbb{R}^3\}_N$ is the discrete average operator. Its definition is

$$\mathcal{A}^h: \{\mathbb{R}^3\}_N \rightarrow \{\mathbb{R}^3\}_N,\ y_{j=1,\dots,N+1} = \mathcal{A}^h\left(x_{i=1,\dots,N}\right) = \begin{cases} \frac{x_1}{2},\ \text{if } j = 1 \\ \frac{x_j + x_{j-1}}{2},\ \text{if } 1 < j \leq N \\ \frac{x_N}{2},\ \text{if } j = N+1 \end{cases}$$

Together, the discretized Euler-Lagrangian equation is

$$\begin{cases} \dfrac{\partial \boldsymbol{r}_i}{\partial t} = \boldsymbol{v}_i,\ i = [1,\ n+1] \\ \dfrac{\partial \boldsymbol{d}_{i,j}}{\partial t} = \left(\boldsymbol{Q}_i^T \boldsymbol{\omega}_{\mathcal{L}}^i\right) \times \boldsymbol{d}_{i,j}, i = [1, n],\ \ j = 1,2,3 \\ m_i \cdot \dfrac{\partial \boldsymbol{v}_i}{\partial t} = \Delta^h \left(\dfrac{\boldsymbol{Q}_i^T \widehat{\boldsymbol{S}}_i \boldsymbol{\sigma}_{\mathcal{L}}^i}{e_i}\right) + \boldsymbol{F}_i, i = [1, n+1] \\ \dfrac{\widehat{\boldsymbol{J}}_i}{e_i} \cdot \dfrac{\partial \boldsymbol{\omega}_{\mathcal{L}}^i}{\partial t} = \Delta^h \left(\dfrac{\widehat{\boldsymbol{B}}_i \widehat{\boldsymbol{\kappa}}_{\mathcal{L}}^i}{{\varepsilon_i}^3}\right) + \mathcal{A}^h \left(\dfrac{\widehat{\boldsymbol{\kappa}}_{\mathcal{L}}^i \times \widehat{\boldsymbol{B}}_i \widehat{\boldsymbol{\kappa}}_{\mathcal{L}}^i}{{\varepsilon_i}^3} l_i\right) + \left(\boldsymbol{Q}_i \boldsymbol{t}_i \times \widehat{\boldsymbol{S}}_i \boldsymbol{\sigma}_{\mathcal{L}}^i\right) \hat{l}_i + \left(\widehat{\boldsymbol{J}}_i \cdot \dfrac{\boldsymbol{\omega}_{\mathcal{L}}^i}{e_i}\right) \times \boldsymbol{\omega}_{\mathcal{L}}^i + \dfrac{\widehat{\boldsymbol{J}}_i \boldsymbol{\omega}_{\mathcal{L}}^i}{e_i^2} \cdot \dfrac{\partial e_i}{\partial t} + \boldsymbol{C}_{\mathcal{L}}^i, i = [1, n] \end{cases}$$

The dynamic simulation of a slender body can be solved computationally with an ordinary differential equation. In our simulation, we used Runge-Kutta 4th order method to numerically integrate the functions over time. Compared to the Kirchhoff description, Lagrange multipliers that enforce the condition of inextensibility and unshearability are no longer needed.

## Two-dimensional simplification of a single Cosserat rod

To simplify the problem and understand the locomotion principle of the snake traversing obstacles with large height variation, we only consider the problem in two dimensions. When simplifying the dynamic function into two dimensional, the rotation matrix is

$$Q = \begin{bmatrix} \cos\theta & \sin\theta & 0 \\ -\sin\theta & \cos\theta & 0 \\ 0 & 0 & 1 \end{bmatrix}$$

Here, the z coordinate is pointing out of the paper. The continuous Eulerian-Lagrangian form becomes:

$$
\begin{cases}
\dfrac{\partial r_x}{\partial t} = v_x \\
\dfrac{\partial r_y}{\partial t} = v_y \\
\dfrac{\partial \theta}{\partial t} = \omega_z \\
\mathrm{d}m \cdot \dfrac{\partial [v_1,\ v_2,\ 0]^{\mathrm{T}}}{\partial t} = \dfrac{\partial}{\partial \hat{s}} \left( \dfrac{\boldsymbol{Q}^{\mathrm{T}} \widehat{\boldsymbol{S}} \boldsymbol{Q} \boldsymbol{\sigma}}{e} \right) \mathrm{d}\hat{s} + \boldsymbol{F} \\
\dfrac{d\hat{J}_3}{e} \cdot \dfrac{\partial \omega_z}{\partial t} = \dfrac{\partial}{\partial \hat{s}} \left( \dfrac{B_3}{e^3} \cdot \dfrac{\partial \theta}{\partial \hat{s}} \right) d\hat{s} + \boldsymbol{Q}\big(\boldsymbol{t} \times (\boldsymbol{Q}^{\boldsymbol{T}} \widehat{\boldsymbol{S}} \boldsymbol{Q}) \boldsymbol{\sigma}\big)|_{\boldsymbol{d}_3} d\hat{s} + \dfrac{d\hat{J}_3}{e^2} \cdot \omega_z \cdot \dfrac{\partial e}{\partial t} + C_{\mathcal{L}}
\end{cases}
$$

The shearing rigidity matrix $S$ in the lab frame is

$$
\boldsymbol{S} = \boldsymbol{Q}^{\mathrm{T}} \widehat{\boldsymbol{S}} \boldsymbol{Q} = \begin{bmatrix} \hat{S}_1 \cos^2\theta + S_2 \sin^2\theta & (\hat{S}_1 - \hat{S}_2) \sin\theta \cos\theta & 0 \\ (\hat{S}_1 - \hat{S}_2)\, sin\, \theta\, cos\, \theta & \hat{S}_1 \sin^2\theta + \hat{S}_2 \cos^2\theta & 0 \\ 0 & 0 & \hat{S}_3 \end{bmatrix} = \begin{bmatrix} S_{11} & S_{12} & 0 \\ S_{12} & S_{22} & 0 \\ 0 & 0 & S_{33} \end{bmatrix}
$$

Then we can obtain the discretized Eulerian-Lagrangian form in the way described before

$$
\begin{cases}
\dfrac{\partial r_{x,i}}{\partial t} = v_{x,i},\ i = [1,\ n+1] \\
\dfrac{\partial r_{y,i}}{\partial t} = v_{y,i},\ i = [1,\ n+1] \\
\dfrac{\partial \theta_i}{\partial t} = \omega_i,\ i = [1,\ n] \\
m_i \dfrac{\partial v_{x,i}}{\partial t} = \Delta_h(f_{int,x,i}) + \mathrm{d}F_{x,i}, i = [1,\ n+1] \\
m_i \dfrac{\partial v_{y,i}}{\partial t} = \Delta_h(f_{int,y,i}) + \mathrm{d}F_{y,i}, i = [1,\ n+1] \\
\dfrac{\mathrm{d}\hat{J}_{3,i}}{e_i} \cdot \dfrac{\partial \omega_{z,i}}{\partial t} = \Delta_h \left( \dfrac{\hat{B}_3}{\varepsilon_{i+1}^3} \cdot \dfrac{\theta_{i+1} - \theta_i}{\hat{l}_i} \right) + (r_{x,i+1} - r_{x,i}) f_{int,y,i} - (r_{y,i+1} - r_{y,i}) f_{int,x,i} + \dfrac{\mathrm{d}\hat{J}_{3,i} \omega_i}{e_i^2} \cdot \dfrac{\partial e_i}{\partial t} + c_i
\end{cases}
$$

Where the internal elastic force in lab frame is

$$
\begin{cases}
f_{int,x,i} = \dfrac{S_{11,i} \sigma_{x,i} + S_{12,i} \sigma_{y,i}}{e_i} \\
f_{int,y,i} = \dfrac{S_{12,i} \sigma_{x,i} + S_{22,i} \sigma_{y,i}}{e_i}
\end{cases}
$$

And the time derivative of the compression or stretching ratio $e$ is

$$\frac{\partial e_i}{\partial t} = \frac{1}{e_i l_i^2}\left[\left(r_{x,i+1} - r_{x,i}\right)\left(v_{x,i+1} - v_{x,i}\right) + \left(r_{y,i+1} - r_{y,i}\right)\left(v_{y,i+1} - v_{y,i}\right)\right]$$

With this, we can use the numerical integration method to compute the evolution of the snake's dynamics. The snake does not stretch or twist at an observable scale, according to the biological findings. Hence, we increased the stretching rigidity and twisting rigidity by order of magnitude to reduce the deformations of corresponding degrees of freedom. With an appropriate young's modulus, the snake's axial and shear strain is small. In this situation, the Cosserat is approaching the Kirchhoff rod, but the Lagrangian multipliers are avoided. This method is a kind of penalty method. It is noted that the time step should be low enough to prevent overshooting, which may lead to the failure of convergence.

## Shape control and gait generation

The snake-terrain interaction defines the external forces exerted on the snake. The normal force is modeled by the mass-spring-damper model. The damping coefficient is set as the critical value to prevent bouncing back or oscillation after a collision. The normal repulsive force is written as

$$F_N = \begin{cases} k_s s + k_v \dot{s}, s \geq 0 \\ 0, s < 0 \end{cases}$$

Where s is the penetration, the normal repulsive force exists only when the penetration is positive. Suppose the mass of the element is m, and the natural frequency is set as $\omega_n$. To make the critically damping

$$k_s = m\omega_n^2, k_s = 2m\omega_n^2$$

The stiffness of the terrain can be tuned accordingly to adjust the compliance of the terrain. As for the tangential direction, a Coulomb's friction is exerted on the snake. The snake's body is set as homogeneous and with a constant radius to simplify the problem.

The actuator mimicking the snake's muscular force is used as the internal torque. In order to achieve the prescribed shape, a shape feedback controller is designed to get the value of the internal torque. The angular acceleration of the slender body is linearly dependent on the internal torque, with the following form:

$$J_i \ddot{\theta}_i = N_i + C_i$$

Where $N_i$ is the torque due to deformation, and $C_i$ is the torque given by the actuator. Let the desired shape angle be $\theta_{d,i}$ ($i = 1, \ldots, n$), we can propose the shape feedback controller

$$\ddot{\theta}_i = \ddot{\theta}_{d,i} + k_1(\dot{\theta}_{d,i} - \dot{\theta}_i) + k_0(\theta_{d,i} - \theta_i)$$

When the gain $k_1$ and $k_0$ are positive, the close loop is stable. If the natural frequency is $\omega_n$, let $k_1 = 2\omega_n$ and $k_0 = \omega_n^2$ to make the close loop critically damped.

The initial shape of the snake is set as the natural conforming to an isosceles triangle. Part of the left section of the isosceles triangle overlaps the triangular wedge. The inflection of the snake shape is slightly higher than the wedge's highest vertices to avoid collisions. The pure propagation gait is a control strategy to shift the shape of the anterior section down to the posterior section. In our case, the terrain ahead of the snake is flat. Hence, the shape angle will all decay to 0 at the end.

## Force analysis based on a simplified model

Using a simplified model, we can derive the snake-terrain interaction. The snake can be divided into three sections in contact with the terrain, as shown in Fig. 4. Another dangling section is not in contact with the terrain. We assume that the snake follows a tube-like motion without lateral slip, and the tangential speed magnitude is $v_t$. The momentum of the snake is

$$\begin{cases} p_x = m v_t \dfrac{l_x}{l} \\ p_z = m v_t \dfrac{l_z}{l} \end{cases}$$

Where $l_x$ and $l_y$ are the horizontal and vertical projections of the snake shape curve. We assume that in our case, the horizontal projection is $l_x = l[1 - 2\lambda(1 - \cos\theta)]$, where $\lambda$ is the ratio between the isosceles edge and the body length. While the head and tail lay on the same level in terms of vertical coordinate, yielding $l_z = 0$. Take the time derivative of the momentum

$$\begin{cases} \dfrac{dp_x}{dt} = ma_t[1 - 2\lambda(1 - \cos\theta)] \\ \dfrac{dp_z}{dt} = 0 \end{cases}$$

Where $a_t$ is the magnitude of the tangential acceleration. According to Newton's second law

$$\begin{cases} \dfrac{dp_x}{dt} = -\mu(F_{N1} + F_{N4}) + F_{N3}\sin\theta - \mu F_{N3}\cos\theta \\ \dfrac{dp_z}{dt} = F_{N1} + F_{N4} + F_{N3}\cos\theta + \mu F_{N3}\sin\theta - mg \end{cases}$$

The friction force is chosen as the kinematic friction subject to Coulomb's law of friction. Solve the above equations, we got

$$\begin{cases} F_{N1} + F_{N4} = \dfrac{1 - \mu\cot\theta}{1 + \mu^2} mg - \dfrac{\mu + \cot\theta}{\mu^2 + 1} ma_t[1 - 2\lambda(1 - \cos\theta)] \\ F_{N3} = \dfrac{\mu}{(\mu^2 + 1)\sin\theta} mg + \dfrac{ma_t[1 - 2\lambda(1 - \cos\theta)]}{(\mu^2 + 1)\sin\theta} \end{cases}$$

Apparently, when the snake accelerates, the reaction force on the slope increases and the reaction force on the flat ground decreases. In the steady state where the acceleration equals zero, the normal force becomes

$$\begin{cases} F_{N1} + F_{N4} = \dfrac{1 - \mu\cot\theta}{1 + \mu^2} mg \\ F_{N3} = \dfrac{\mu}{(\mu^2 + 1)\sin\theta} mg \end{cases}$$

Note that we only used force balance and the directions of $F_{N1}$ and $F_{N3}$ are identical, so we can only obtain the sum of $F_{N1}$ and $F_{N3}$ here. If in a steady state, the repulsive force on the slope remains constant during the snake's traversal of the wedge. In order to make the normal repulsive positive, the wedge slope $\tan\theta$ must be greater than the friction coefficient, which is a precondition in our problem. When $\tan\theta$ equals the friction coefficient, the normal force on the flat ground equals zero and the entire body weight is supported by the slope. The propulsive force on the slope also increases with the friction coefficient $\mu$ and decreases with the slope angle $\theta$, suggesting that a large slope and small friction coefficient benefit the traversal. On the other hand, as $F_{N3}$ remains constant with fixed slope angle and friction coefficient, the pressure on the slope decreases with the contact area. We can make a hypothesis that a large wedge has a large contact area for propulsion that benefits the traversal.

Using the simplified model, the position of the center of mass is

$$\begin{cases} x_c/l = [0.5 - \lambda(1 - \cos\theta)] + 2\lambda(1 - 2\lambda)(1 - \cos\theta)(0.5 - k) \\ z_c/l = \lambda^2 \sin\theta \end{cases}$$

The angular momentum with respect to the center of mass is

$$L = \lambda^2 \sin\theta\,[2(1 - \cos\theta)(1 - \lambda) - 1] m v_t l$$

Take the derivative

$$\frac{dL}{dt} = \lambda^2 \sin\theta\,[2(1 - \cos\theta)(1 - \lambda) - 1] m a_t l$$

The overall external torque equals the derivative of angular momentum according to the angular momentum theorem. If the motion is steady, the angular momentum is time-invariant, which means the overall external torque with respective the center of mass equals zero. We can compute the geometrical relationship of force application points based on torque balance. We assume that the point of force application on the slope is at the middle of the section. This refers to the situation in which the terrain reaction force distributes evenly on the slope. Since the terrain reaction forces on the flat ground share the same direction, we can combine $F_1$ and $F_4$ to a joint force, shown in Fig. 5. There are three forces in total, in order to satisfy torque balance, the three forces have to intersect at one point. The points of force application of the gravitational force $G$ and the terrain reaction force on the slope $F_3$ are known. Hence, the intersection point is fixed. The joint force of $F_1$ and $F_4$ must pass through the intersection. Since the points of application of $F_1$ and $F_4$ must be on the ground level, the point of force application of their joint force is determined at the $[x_{14} \quad 0]^T$ shown in the figure. The next step is to determine the separate points of force application of $F_1$ and $F_4$, which can be decomposed as

$$(F_1 + F_4)x_{14} = F_1 x_1 + F_4 x_4$$

As the ratio of $F_1$ and $F_4$ is unknown, the equation is redundant. The points of force application of $F_1$ and $F_4$ should be in the range of each section, which can be used to determine if the snake is eligible for traversing in specific circumstances.

## Internal force

We can derive the internal force and torque based on the same simplified model. Again, we suppose the snake's tangential speed is $v_t$ and acceleration is $a_t$. We conduct a free body diagram analysis of an infinitesimal straight section, shown in Fig. 5. Internal forces include the tension and shearing forces, and external forces include the gravitational force and the terrain reaction force.

$$\begin{cases} \lambda_m \mathrm{d}s \cdot a_t = \mathrm{d}T - \lambda_m g \mathrm{d}s \cdot \sin\alpha - \mu \mathrm{d}F_N \\ 0 = \mathrm{d}S - \lambda_m g \mathrm{d}s \cdot \cos\alpha + \mathrm{d}F_N \end{cases}$$

Where $\lambda_m$ is the linear density of the snake, $\alpha$ is the inclination angle. Rearrange the equations, we have

$$\begin{cases} \dfrac{\mathrm{d}T}{\mathrm{d}s} = \lambda_m a_t + \lambda_m g \sin\alpha + \mu \dfrac{\mathrm{d}F_N}{\mathrm{d}s} \\ \dfrac{\mathrm{d}S}{\mathrm{d}s} = \lambda_m g \cos\alpha - \dfrac{\mathrm{d}F_N}{\mathrm{d}s} \end{cases}$$

Apply to the different sections we have

$$\text{Section 1: } \begin{cases} \frac{\mathrm{d}T}{\mathrm{d}s} = \lambda_m a_t + \mu \frac{\mathrm{d}F_N}{\mathrm{d}s} \\ \frac{\mathrm{d}S}{\mathrm{d}s} = \lambda_m g - \frac{\mathrm{d}F_N}{\mathrm{d}s} \end{cases}$$

$$\text{Section 2: } \begin{cases} \frac{\mathrm{d}T}{\mathrm{d}s} = \lambda_m a_t + \lambda_m g \sin\theta \\ \frac{\mathrm{d}S}{\mathrm{d}s} = \lambda_m g \cos\theta \end{cases}$$

$$\text{Section 3: } \begin{cases} \frac{\mathrm{d}T}{\mathrm{d}s} = \lambda_m a_t - \lambda_m g \sin\theta + \mu \frac{\mathrm{d}F_N}{\mathrm{d}s} \\ \frac{\mathrm{d}S}{\mathrm{d}s} = \lambda_m g \cos\theta - \frac{\mathrm{d}F_N}{\mathrm{d}s} \end{cases}$$

$$\text{Section 4: } \begin{cases} \frac{\mathrm{d}T}{\mathrm{d}s} = \lambda_m a_t + \mu \frac{\mathrm{d}F_N}{\mathrm{d}s} \\ \frac{\mathrm{d}S}{\mathrm{d}s} = \lambda_m g - \frac{\mathrm{d}F_N}{\mathrm{d}s} \end{cases}$$

Note that section 2 is dangling, resulting in zero normal force. The internal forces on the inflection section are different from the straight section. We conduct a free body diagram analysis of the inflection section shown in Fig. 5. The inflection angle is $\theta$, the length of the flat section and the inclined section is $\mathrm{d}s_1$ and $\mathrm{d}s_2$, respectively. The linear momentum is

$$\begin{cases} p_x = \lambda_m v_t (\mathrm{d}s_1 + \mathrm{d}s_2 \cos\theta) \\ p_y = \lambda_m v_t \mathrm{d}s_2 \sin\theta \end{cases}$$

Following the angle shifting strategy, we can take the time derivatives of $\mathrm{d}s_1$ and $\mathrm{d}s_2$ assuming the forward motion is towards the right

$$\mathrm{d}\dot{s}_1 = -v_t$$

$$\mathrm{d}\dot{s}_2 = v_t$$

Then we can take the time derivative of the linear momentum, which gives

$$\begin{cases} \dot{p}_x = \lambda_m a_t (\mathrm{d}s_1 + \mathrm{d}s_2 \cos\theta) - \lambda_m v_t^2 (1 - \cos\theta) \\ \dot{p}_y = \lambda_m a_t \mathrm{d}s_2 \sin\theta + \lambda_m v_t^2 \sin\theta \end{cases}$$

When $\mathrm{d}s_1$ and $\mathrm{d}s_2$ approaches zero, the derivative of the linear momentum is close to

$$\begin{cases} \dot{p}_x = -\lambda_m v_t^2 (1 - \cos\theta) \\ \dot{p}_y = \lambda_m v_t^2 \sin\theta \end{cases}$$

The forces exerted on the inflection section consist of the tension on both sides $T_L$ and $T_R$, the shearing on both sides $S_L$ and $S_R$, and the gravitational force $G$. According to Newton's second law,

$$\begin{cases} \dot{p}_x = -T_L + T_R \cos\theta - S_R \sin\theta \\ \dot{p}_y = -S_L + T_R \sin\theta + S_R \cos\theta - G \end{cases}$$

If the internal forces on the left side are given, we can solve for the right limit of the internal forces

$$\begin{cases} T_{1R} = (\dot{p}_x + T_{1L}) \cos\theta + \left(\dot{p}_y + G + S_{1L}\right) \sin\theta \\ S_{1R} = -(\dot{p}_x + T_{1L}) \sin\theta + \left(\dot{p}_y + G + S_{1L}\right) \cos\theta \end{cases}$$

Assume that $\mathrm{d}s_1$ and $\mathrm{d}s_2$ approaches zero, the gravitational force is negligible. Plugging the momentum rate into the equations yields

$$\begin{cases} T_{1R} = \lambda_m v_t^2 (1 - \cos\theta) + T_{1L} \cos\theta + S_{1L} \sin\theta \\ S_{1R} = \lambda_m v_t^2 \sin\theta - T_{1L} \sin\theta + S_{1L} \cos\theta \end{cases}$$

Which shows the discrepancy of internal forces on the inflection point. Similarly, we can derive the condition in the second inflection point

$$\begin{cases} T_{2R} = \lambda_m v_t^2 (1 - \cos 2\theta) + T_{2L} \cos 2\theta - S_{2L} \sin 2\theta \\ S_{2R} = -\lambda_m v_t^2 \sin 2\theta + T_{2L} \sin 2\theta + S_{2L} \cos 2\theta \end{cases}$$

And the third inflection point

$$\begin{cases} T_{3R} = \lambda_m v_t^2 (1 - \cos\theta) + T_{3L} \cos\theta + S_{3L} \sin\theta \\ S_{3R} = \lambda_m v_t^2 \sin\theta - T_{3L} \sin\theta + S_{3L} \cos\theta \end{cases}$$

Finally, if the ground reaction force is known, then we can derive the expression of the internal forces. By integrating the equation xxx, we can derive the internal forces on the inflection points. Assume the mass of sections 1 to 4 are $m_1$, $m_2$, $m_3$, and $m_4$, respectively. As the internal forces at the end of the snake are zero, integrating the section 1 yields

$$\begin{cases} T_{1L} = m_1 a_t + \mu F_{N1} \\ S_{1L} = m_1 g - F_{N1} \end{cases}$$

Section 2 has the relations

$$\begin{cases} T_{2L} = T_{1R} + m_2 a_t + m_2 g \sin\theta \\ S_{2L} = S_{1R} + m_2 g \cos\theta \end{cases}$$

Section 3 has the relations

$$\begin{cases} T_{3L} = T_{2R} + m_3 a_t - m_3 g \sin\theta + \mu F_{N3} \\ S_{3R} = S_{2R} + m_3 g \cos\theta - F_{N3} \end{cases}$$

Section 4 has the relations

$$\begin{cases} 0 = T_{3R} + m_4 a_t + \mu F_{N4} \\ 0 = S_{3R} + m_4 g - F_{N4} \end{cases}$$

Note that the internal forces eliminate to zero at the end of the snake. Given the discrepancy of the inflection points, we can obtain the internal forces on all those inflection points.

## Internal torque analysis

The internal torque is continuous on the inflection points. If not, we can select an infinitesimal section, and the torque discrepancy will result in an infinite large angular acceleration.

The torque along the snake can be computed by the differential equation

$$\mathrm{d}\tau = 0.5\mathrm{d}m \cdot g\cos\theta - 0.5\mathrm{d}F_N - S\mathrm{d}s$$

The torque computed here can be decomposed into two parts, the driving torque and the elastic bending. The bending only occurs at the deflection section and can be computed by the local curvature. Therefore, if the terrain reaction force is known, we can derive the internal forces which in turn can compute the driving torque.

### Normalization

To compare the dynamics across different magnitudes of parameters, we normalize the result to be dimensionless. The unit of length is chosen as the length of the snake $L$, the mass as the mass of the snake $M$, and the force as the gravity of the snake $Mg$. Hence, the time unit is $\sqrt{L/g}$, the velocity unit is $\sqrt{gL}$, Young's modulus unit is $Mg/L^2$. For instance, equation xxx can be normalized to

$$\begin{cases} \bar{F}_{N1} + \bar{F}_{N4} = \dfrac{1 - \mu \cot\theta}{1 + \mu^2} \\ \bar{F}_{N3} = \dfrac{\mu}{(\mu^2 + 1)\sin\theta} \end{cases}$$

The normalization can let us compare cases across a different magnitude of parameters.

## Results

As a first step to understanding the locomotion of snake traversing obstacles with height variation, we used a setting of a single right triangular block. The snake can only generate propulsion at its hypotenuse except for the condition that the snake moves too fast and collides with the other edge. We first used the pure propagation gait in the simulation. The case shown here holds the terrain friction coefficient 0.2, the wedge height 0.1 m, and the slope 0.5. The snake length is 2 m. We can output the driving torque, terrain reaction force, and internal force, which are all difficult to measure in real animal experiments. We used the pure propagation gait to control the snake to move forward. In the simulation, the snake propagates its body shape backward with a constant speed to achieve a near steady speed forward motion. The motion is intermittent without force feedback control. We found that the torque's maximum is in the middle of the body around the wedge obstacle. Torque and normal force oscillate periodically with the intermittency of the body velocity. Remarkably, the peak normal force linear density on the slope is as large as ten times of normal force linear density on the flat ground. This suggests that snakes hold a large capacity to use vertical bending to push against the environment to generate propulsion.

Fig. 6 and Fig. 7 shows the distribution of the terrain reaction force normal component along the snake body. The result is normalized by the body weight. The terrain reaction force the normal component of sections on the flat ground to either support the body weight or zero, which is also shown in Fig. 8a. We can use this to allocate the ratio of $F_{N1}$ and $F_{N4}$ in addition to torque balance. The terrain reaction force normal component is distributed unevenly on the slope, but we assume that it is distributed evenly in our simplified model. This assumption works well in the internal force analysis due to the length of section 3 is relatively short. Fig. 8b shows the result from the simplified model. The terrain reaction force in section 1 aligns with the simulation result, while the terrain reaction force in section 4 differs from the simulation result. The range of the normal force with nonzero value deviates from the simulation result especially when the snake propagates forward. The reaction force given by the simplified model acts as the base of the internal force analysis shown below.

Fig. 9a shows the tension along the snake body. Positive value refers to stretching, and negative value refers to compression. Tension along section 1 is stretching, whose value increases from the left to the right. The cantilevering section (section 2) is mostly stretching, and the magnitude of stretching increases. There is a discontinuous point at the inflection point at the hump. The tension is almost eliminated to zero in section 3, which is on the slope. There is another discontinuous point at the inflection point 3, and the tension in section 4 is mostly negative, implying the compression. This is in accordance with the fact that the front section experiences pushing force from the back. The result computed from the simplified model is shown in Fig. 9b, which aligns with the simulation result quantitatively. Observations from both the simulation results and the model results indicate that the stretching force peaks at the inflection point 2, and the compression force peaks at the inflection point 3.

Fig. 10 shows the shear along the snake body. Fig. 5 shows the positive direction of shear. The shear is almost zero in section 1. In section 2, the shear increases to the maximum. There are discrepancies at the infection point 2 and inflection point 3. The shear increases in the cantilevering section. In section 3, the shear decreases to zero and holds for a while before decreasing to a negative value. Similar to the tension, the shear also peaks at the inflection point 2 and inflection point 3. A schematic is shown is Fig. 11.

Fig. 12 shows the driving toque along the snake body. A posteriorly propagating internal torque profile with a maximum on body segments around the wedge obstacle is displayed in the colormap. Driving torque around the body segments on the flat ground is near zero. The driving torque peak in section 3 where it almost keeps a constant value. There are discrepancies at inflection point 2 and 3.

## External dynamics

### Friction coefficient

We first tune the friction coefficient of the terrain. The slope of the triangular wedge is set as 0.5. When the terrain is frictionless, the snake lacks an effective way to brake. Hence, after accelerating by pushing against the obstacle, the snake cannot decelerate until it collides with the hump of the obstacle. When the friction coefficient is 0.2, the snake can achieve a near steady speed forward motion. When mu = 0.24, the snake will lift and slip. Its forward speed is intermittent. Starting from mu=0.25, the snake can't get through. With this, we can define the critical friction coefficient $\mu^*$ as the largest friction coefficient for a successful traversal. Higher friction coefficient results in a lower traversal success rate for vertical bending. But if the friction is too low may result in difficulty for the controller to maintain a steady movement, as the smooth case indicates.

The comparison of internal forces with different friction coefficients is shown in Fig. 14. The tension's magnitude is proportional to the friction coefficient. The shear's magnitude is insensitive to the friction coefficient.

## Slope angle of the obstacle

Keeping the height of the triangular wedge constant, we vary the slope angle. By tunning the friction coefficient, we can obtain the largest possible friction coefficient that ensures a successful traversal under a certain slope angle. It is shown that the critical friction coefficient increases with the slope. The critical friction coefficients across different slopes comply with the precondition that the friction coefficient is less than the slope, most around half of the slope, which is shown in Fig. 15. This implies that the snake cannot distribute its entire body weight on the slope. The simulation result confirms our hypothesis that a large slope benefits the traversal.

## Size of the wedge

Keeping the slope angle constant, we vary the size of the wedge. The triangular wedges in different cases are similar. We set the friction coefficient as 0.24. When the wedge height is 0.05 m, the snake gets stuck, failing to traverse the obstacle. When the wedge height increases to 0.10 m, the snake is barely able to traverse, exhibiting a lifting and slipping sequence. However, when the wedge height increases to over 0.15 m, the snake can propagate its body smoothly. We also measure the critical friction coefficient across different wedge sizes. As shown in Fig. 16, the critical friction coefficient increases with the size of the wedge but remains under the slope $\tan\theta$. This suggests that with a large wedge, the snake is more capable of traversing with a higher friction coefficient. This confirms our hypothesis that a large wedge benefits the traversal.

It is intuitive that if the wedge height is zero, the snake lacks a pushing area to propel itself, which gives a zero critical friction coefficient. We tested the case the wedge is extremely small (height = 0.01 m), and it turned out that the propagation of the snake shape ignored the obstacle. When the wedge obstacle is extremely large, we tested the case that the wedge height is 0.35 m. The snake became unstable at the beginning and when the left side left the ground. It is likely because the number of contact surfaces helps maintain stability. When the tail of the snake leaves the flat ground, it is more difficult for the snake to maintain a smooth motion and the cantilevering section is likely to oscillate.

### Stiffness of the terrain

The stiffness of the terrain can be tuned by setting the natural frequency. The default natural frequency is set as 200 Hz. The corresponding time cycle is 0.005 s, which is much larger than the time step $10^{-5}$ s used in our numerical integration. The more terrain stiffness is the more terrain reaction force with the same penetration. As shown in Fig. 17, the intermittency frequency of the snake motion is proportional to the natural frequency of the terrain. This suggests that the stiffness of the terrain affects the snake's motion. The stiffer the terrain is, the more likely the pure propagation gait leads to vertical oscillation with higher frequency. Statistically, higher stiffness with higher damping can eliminate the oscillation of the speed.

### Velocity of the snake

The driving torque and the internal forces are insensitive to the locomotion speed when the speed is relatively slow. The dimensionless speed value is around 0.01 in our cases, which affects little from the perspective of dynamics. Simulations with very high speed have not been conducted.

### Location of the obstacle

We found that the force distribution of the terrain reaction force is important when tunning the parameters. Obviously, it is hard for the snake to distribute its entire body weight on the slope to generate the largest propulsion. The relative location between the snake and the obstacle may result in different situations of force distribution. To test if the relative location of the obstacle affects the traversal, we settle the snake to different locations and then initiate the pure propagation motion. The friction coefficient is set as 0.28 for all cases. We can classify the traversal process into three stages: the beginning, middle, and final stages. For the beginning stage and the final stage, the snake gets stuck. While for the middle stage, the snake can move forward, implying that the snake can generate more propulsion against the wedge to move forward at the middle stage. Hence, if the snake is at the beginning and the final stage, vertical bending alone is not sufficient for propulsion. Other gaits such as concertina are needed for help, which is identical to the experimental observations.

We also tested the case that the snake initiates its pure propagation motion at the critical point between the beginning stage and the middle stage. The snake finally gets stuck at the critical point between the middle stage and the final stage. This suggests that vertical bending alone may only apply to the middle stage. We also tuned the friction coefficient and summarized the stage range. Case of $\mu = 0.28$ is boxed in the colormap. The range of the middle stage becomes narrow as the friction coefficient increases. The range of the beginning stage is greater than the range of the final stage.

The reason why the snake can generate more propulsion against the obstacle at the middle stage is due to the torque balance. We can show this relationship in a theoretic model. Terrain reaction force must be positive. However, in order to satisfy the torque balance, the beginning stage and the final stage give negative terrain reaction force, which is shown in Fig. 19. Cases in which the point of force application lie beyond the corresponding physical section, are also excluded from the possible scenario. Based on this, the model can roughly speculate the range of the middle stage. The simulation result gives the middle stage of $k = [0.36, 0.86]$, while the theoretical model gives the middle stage of $k = [0.30, 0.84]$. The closer the center of mass is to the obstacle, the more likely the snake can generate more propulsion against the obstacle. This can be applied to cases with more than one pushing anchor.

## Internal dynamics

## Radius of the snake

The radius of the snake is much smaller than its length under the assumption of the rod. The value of radius affects the magnitude of torque given by the friction force. However, it is relatively small and is not observable in the simulation results.

## Selection of Young's modulus

The value of Young's modulus is selected by the inspection of varying the value. The higher Young's modulus gives rise to a more rigid rod, requiring higher internal torque actuation. Further, the difference between the cross-section normal direction and the centerline tangential direction is proportional to the shearing force. And this results in a larger deviation between the cross-section normal direction and the centerline tangential direction. To eliminate such effects, we increase the shearing rigidity by order of magnitude. The Young's modulus is set as $10^5$ Pa unless otherwise specified.

## Force feedback control

With pure propagation, the variation of the snake's speed is large. The snake is always experiencing an over-pushing and braking sequence. In a typical case of desired speed equaling 0.06 m/s, the snake ends up with its speed oscillating between 0.03 m/s and 0.09 m/s under pure propagation. Hence, we seek a way to stabilize its motion. The control target is tracking at a steady speed. The only source for the snake to gain propulsion is through the slope. Hence managing the force on the slope can be helpful. To keep the propulsion close to the theoretic value of $F_{N3}$ in a steady motion, the snake can move forward at a near-constant speed after a short period. We can conclude that force feedback helps to control the snake's motion.

However, how to design a proper controller for the snake to perform vertical bending in more complicated terrain remains to be a problem.

# Discussions

## Contribution and implications

Our work utilized a continuous model based on an elastic rod to elucidate the vertical bending mode of snake locomotion. Different from the conventional multi-link rigid model, this model considers the elasticity of the snake and can be actuated by the internal torque. Without the necessity to compute the Lagrange multipliers, the Cosserat model provides a computationally efficient way to simulate the dynamics. As an initial step to reveal the dynamics of snakes using such a locomotor pattern, we found that lower friction and higher slope benefit the traversal. Different from lateral bending, which needs to exploit friction to facilitate movement, vertical bending can take advantage of uneven and even smooth terrain to generate propulsive force. It is potential that gravity pull helps snakes to overcome the friction. As gravity could play an important role in its motion, the relative location of the snake and the obstacle is adjusted, and we found that it is easier to generate propulsion if the center of mass is close to the supporting surface. The efficiency and convenience of simulation can give a good reference to animal experiments in which many data are difficult to collect. Therefore, this method can also be used as a tool to help investigate the snake's other locomotors.

## Limitations and future work

## Relation between the analytic model and the simulation results

The analytic model assumes that the reaction force on the slope is close to an evenly distributed situation, while the reaction force on the slope distributes unevenly in simulation results. The force linear density peaks at the two sides of the slope in the simulation. However, due to the section on the slope being relatively small in length compared to other sections, the internal force analysis accuracy is not affected too much. The trend of the internal force and torque can be the result of the analytic model.

### More complicated scenarios

The simulations conducted only show the case of a snake traversing one obstacle. As a fundamental work to understand the vertical bending, the simple obstacle with fixed surface direction to support the snake elucidates basic rules in vertical bending. However, snakes in the real world tend to use multi obstacles to facilitate their locomotion. An initial study (Fu et al., 2022) showed that snakes could combine lateral and vertical bending in more complicated scenarios. While the effects of terrain feature tendency hold, the effect of multi obstacles may help the snake to maintain its stability which is unclear under current simplified simulations.

### Comparison between the elastic rod model and discrete model

Multi-link model is also widely used to model slender bodies. We also developed a discrete model analogizing the snake robot. The major difference between the discrete rigid model and the continuous elastic model is the intrinsic elasticity. The interactive torque between two adjacent sections in the elastic rod model composes the bending torque and the driving torque. If the number of links is set to be the value of the real robot, the comparison of the continuous model and the discrete model can reveal the difference between the real animal and its robot counterparts. If the number of discrete segments increases to a large value, the two models are similar quantitively since the stiffness of the snake is not large enough to affect the bending torque. This suggests that the results from a continuous model can shed light on the discrete robot model. However, the difference between the animal and the robot is that the animal can dynamically alter body stiffness and the radius of the animal is uneven. Further work is necessary to distinguish these differences.

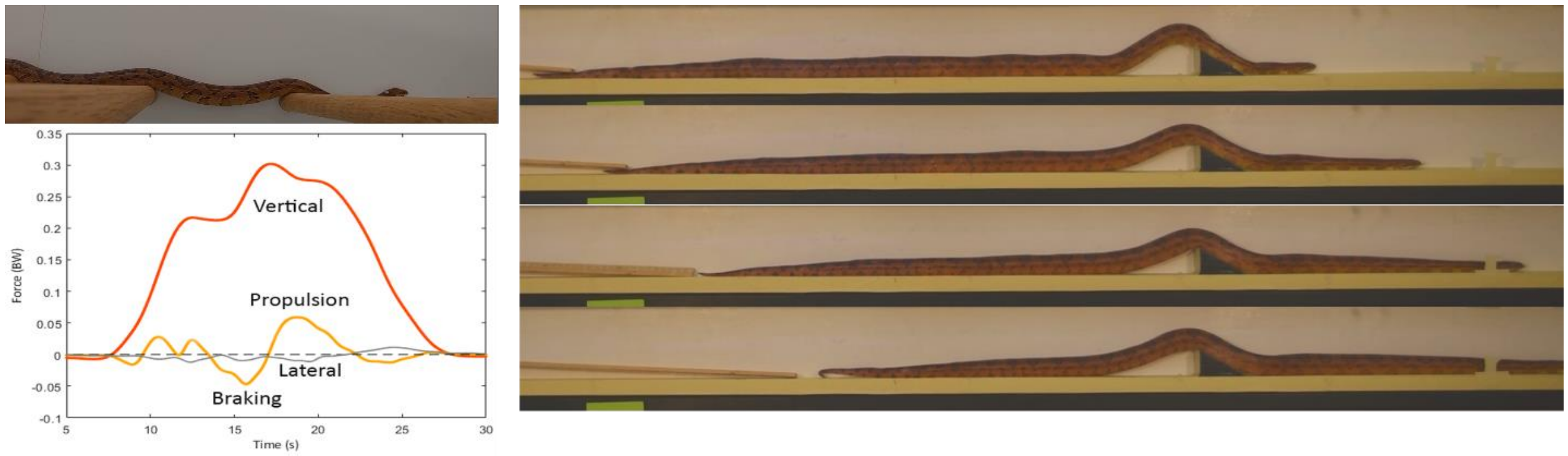

Fig. 1. Animal observations (A) Vertical (red), fore-aft (orange), and lateral (gray solid) forces during a vertical undulation on a horizontal ladder. (B) Side view of a snake traversing a wedge using vertical bending. The snake transitions from a concertina gait to using vertical bending only once it has sufficient contact with the wedge.

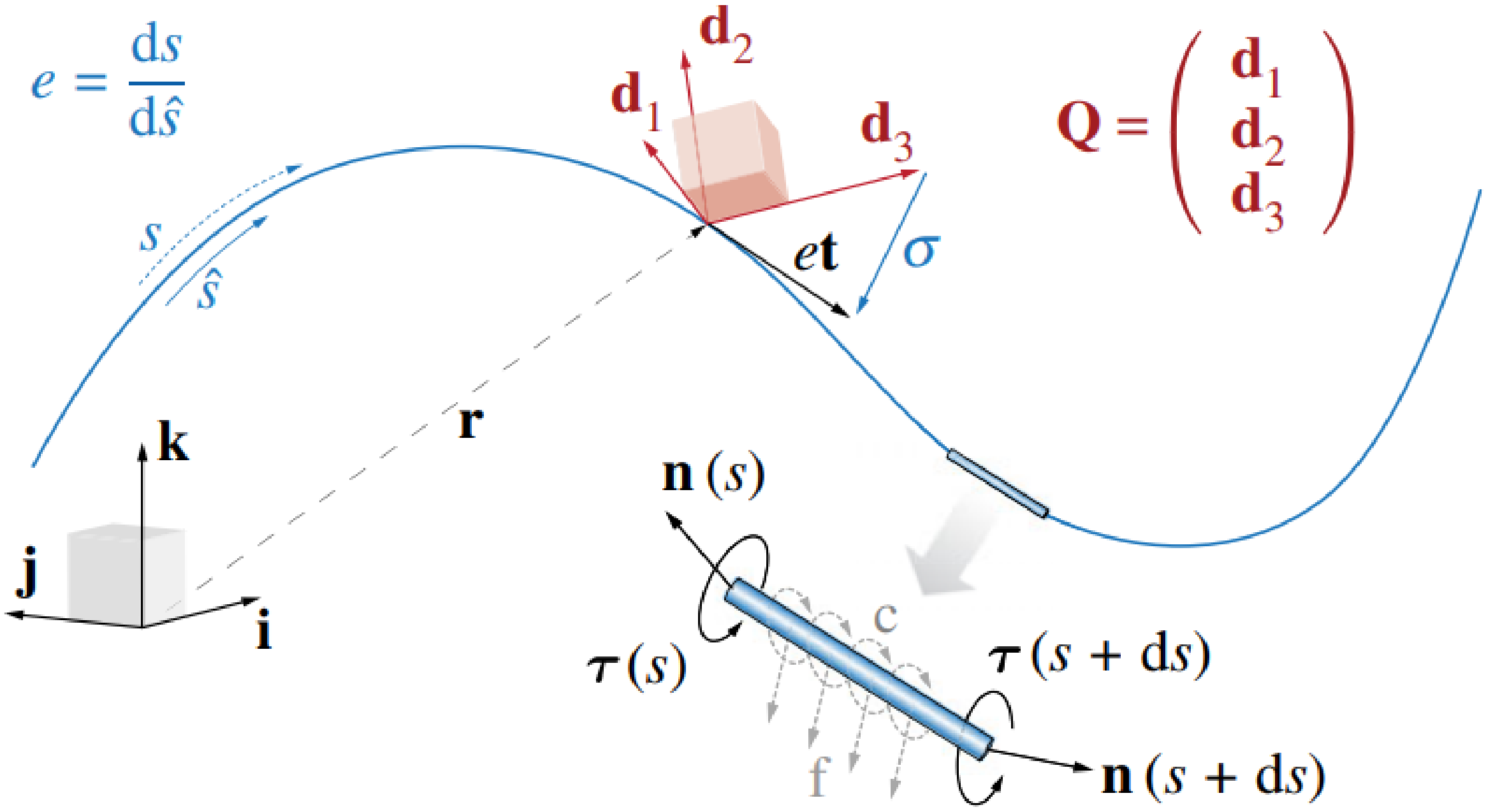

Fig.2 The Cosserat rod model. Adapted from (Gazzola et al., 2018). The rod deforming in the three-dimensonal space is represented by a center-line and a material frame.

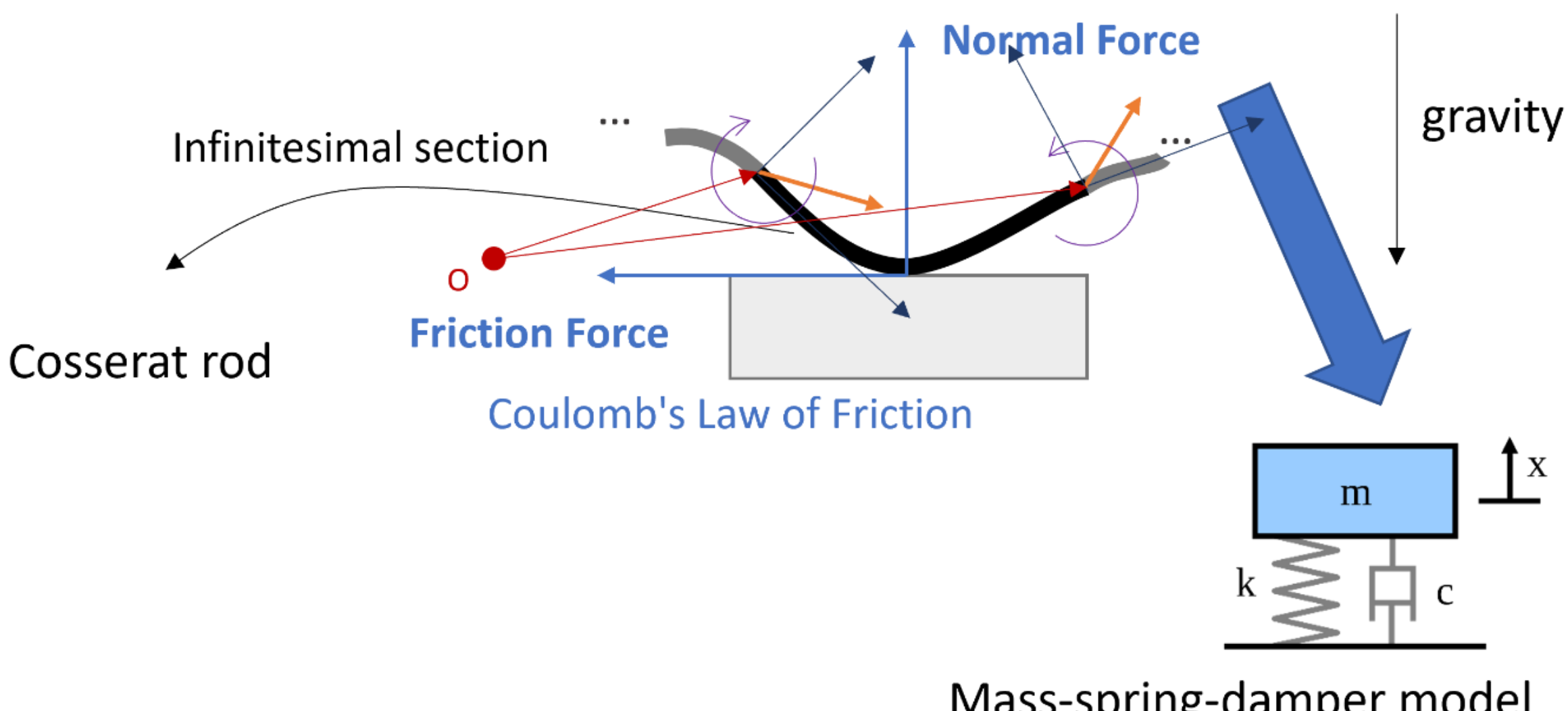

Fig. 3. The external forces include the normal repulsive force which subjects to the mass-spring-damper model, the tangential friction force that follows the Coulomb's law of friction, and the gravitational force.

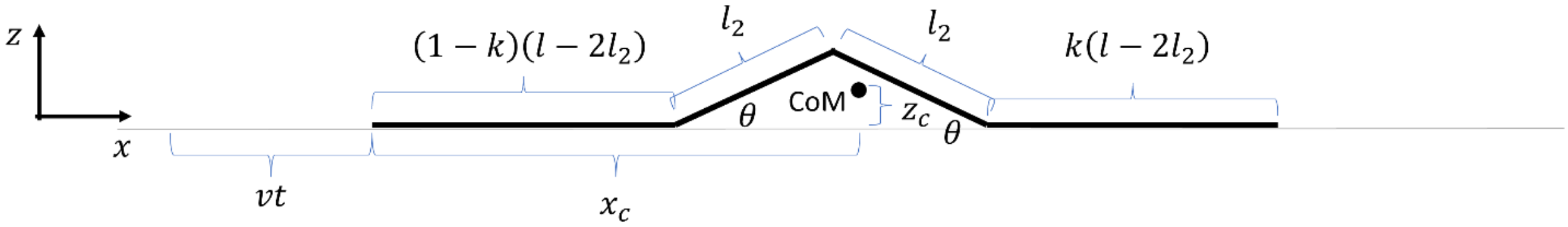

Fig. 4 The geometry of the simplified model. The snake can be divided into three sections in contact with the terrain.

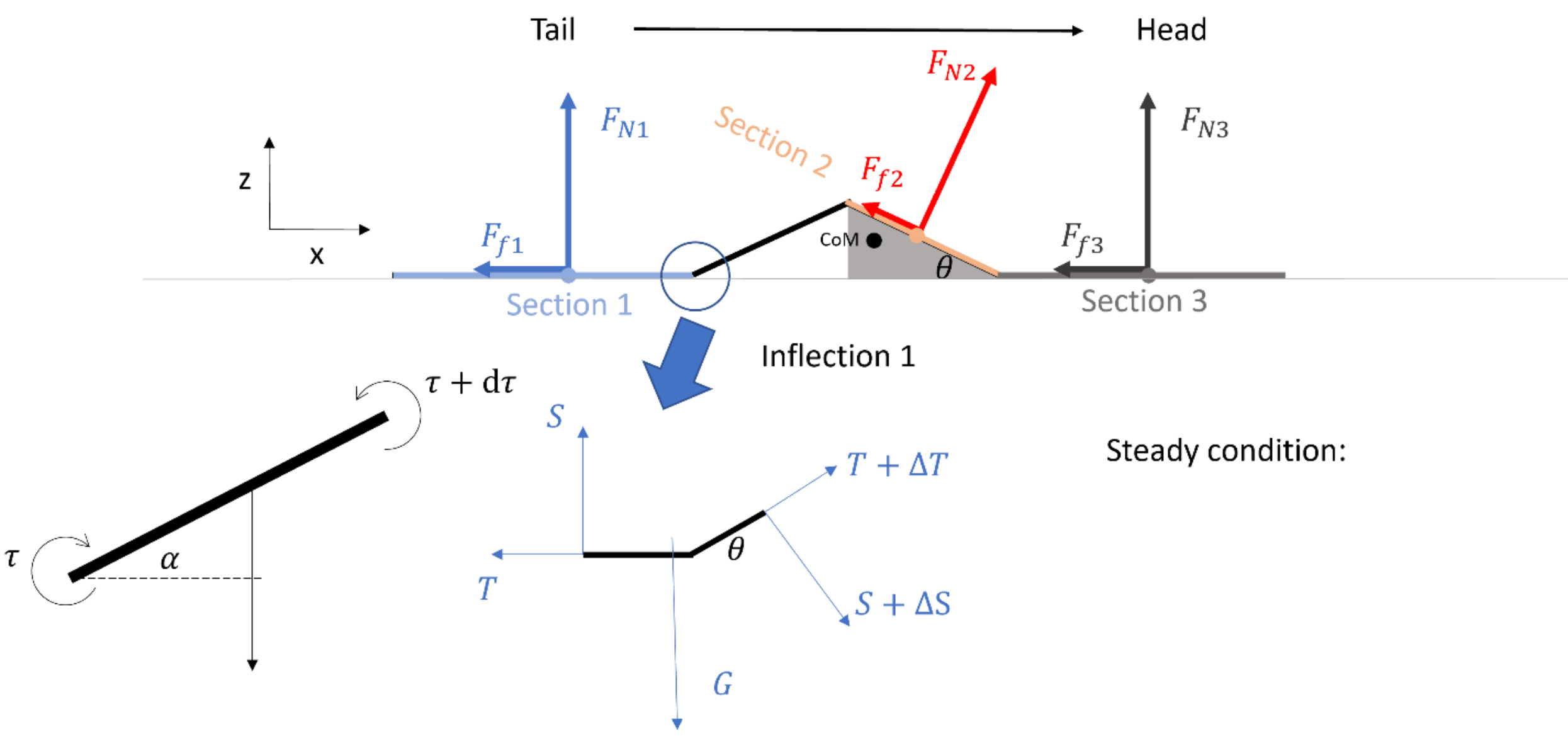

Fig. 5. Force analysis with a simplified model. The snake can be divided into three sections in contact with the ground. The internal force can be obtained through force balance.

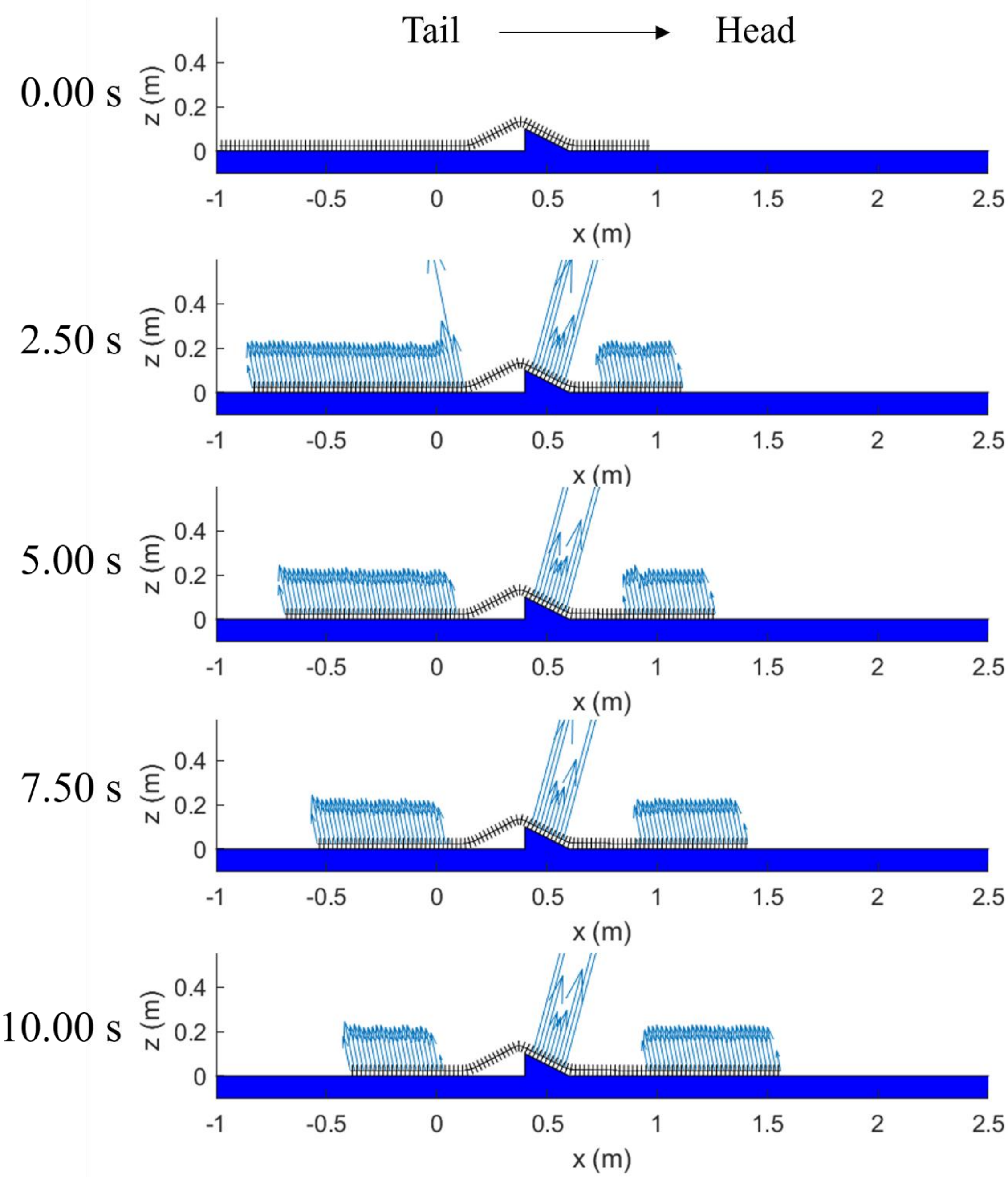

Fig. 6. Evolution of pure propagation with terrain reaction forces annotated.

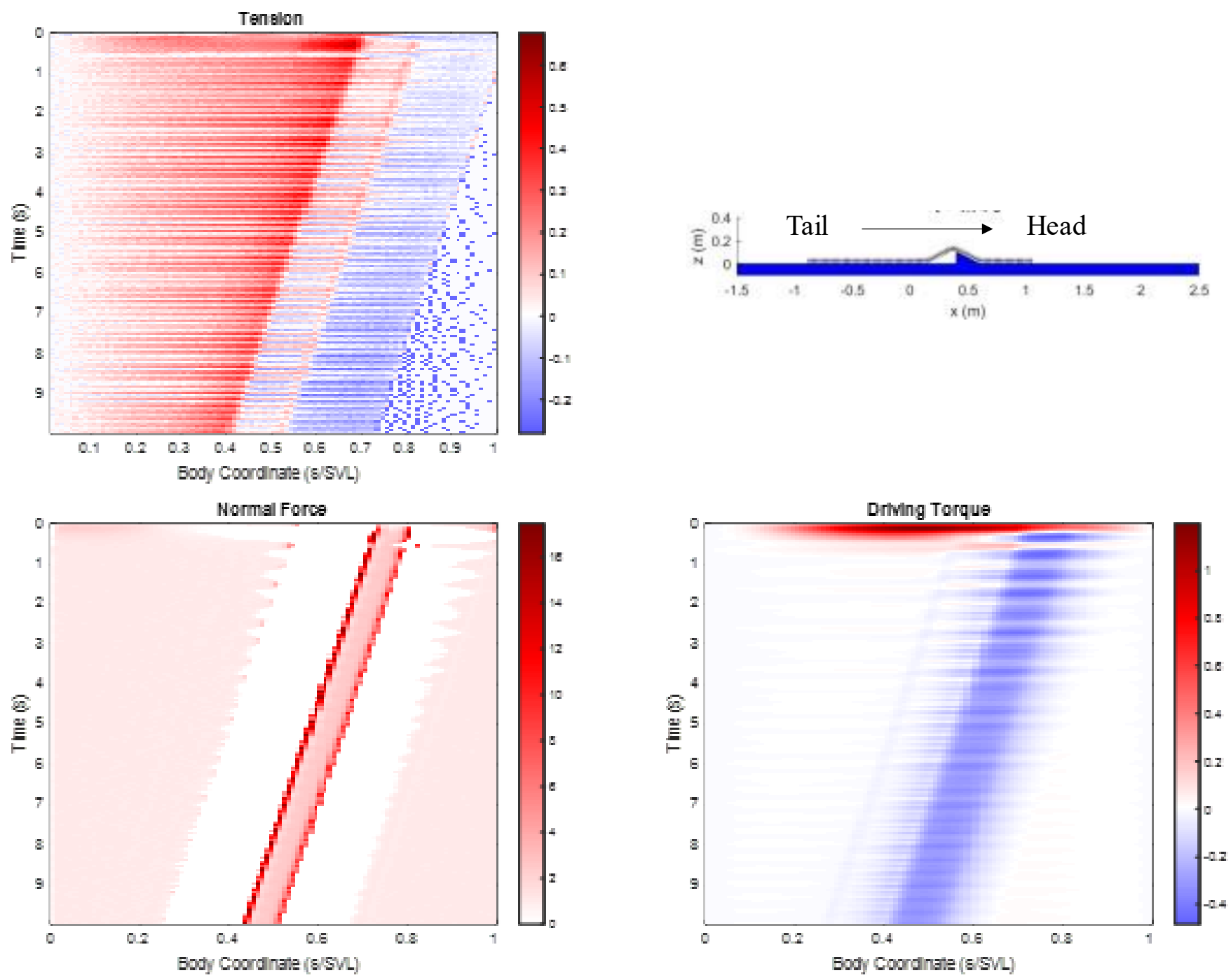

Fig. 7. Torque of the tension, shear, normal force, and driving torque. The tension is positive at the left side, negative at the right side. The shear is zero at the flat ground, negative at the left edge of the hump, positive at the right edge of the hump. The normal force maximizes on the slope. For the part on the ground, the normal force is either equal to its body weight linear density, or zero. The driving torque is zero on the two sides and maximizes around the hump.

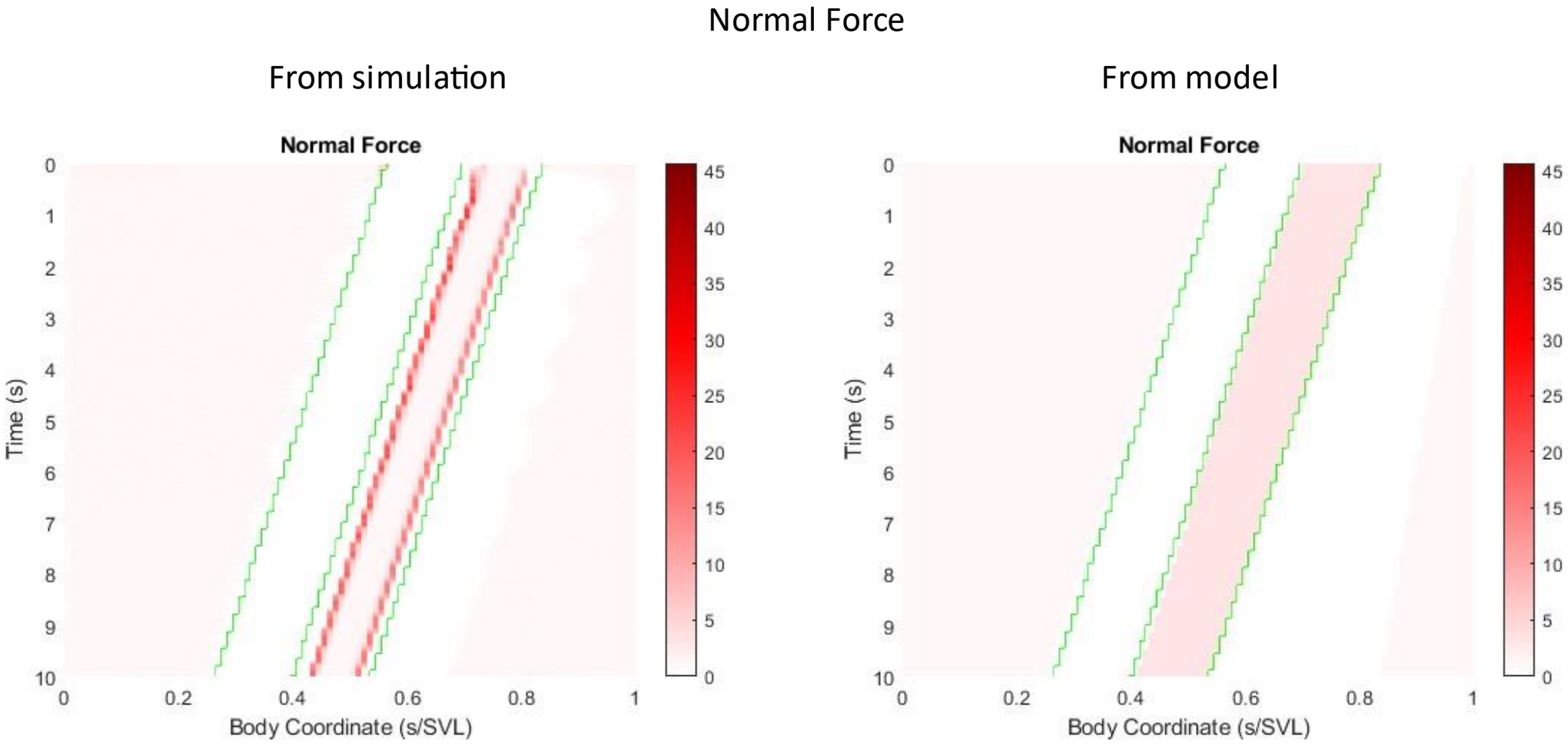

Fig. 8 The distribution of the terrain reaction force normal component along the snake body. The result is normalized by the body weight. The terrain reaction force the normal component of sections on the flat ground to either support the body weight or zero. The model assumes that the normal force distributes evenly on the slope, while the simulation results suggest that the normal force peaks at two sides of the slope.

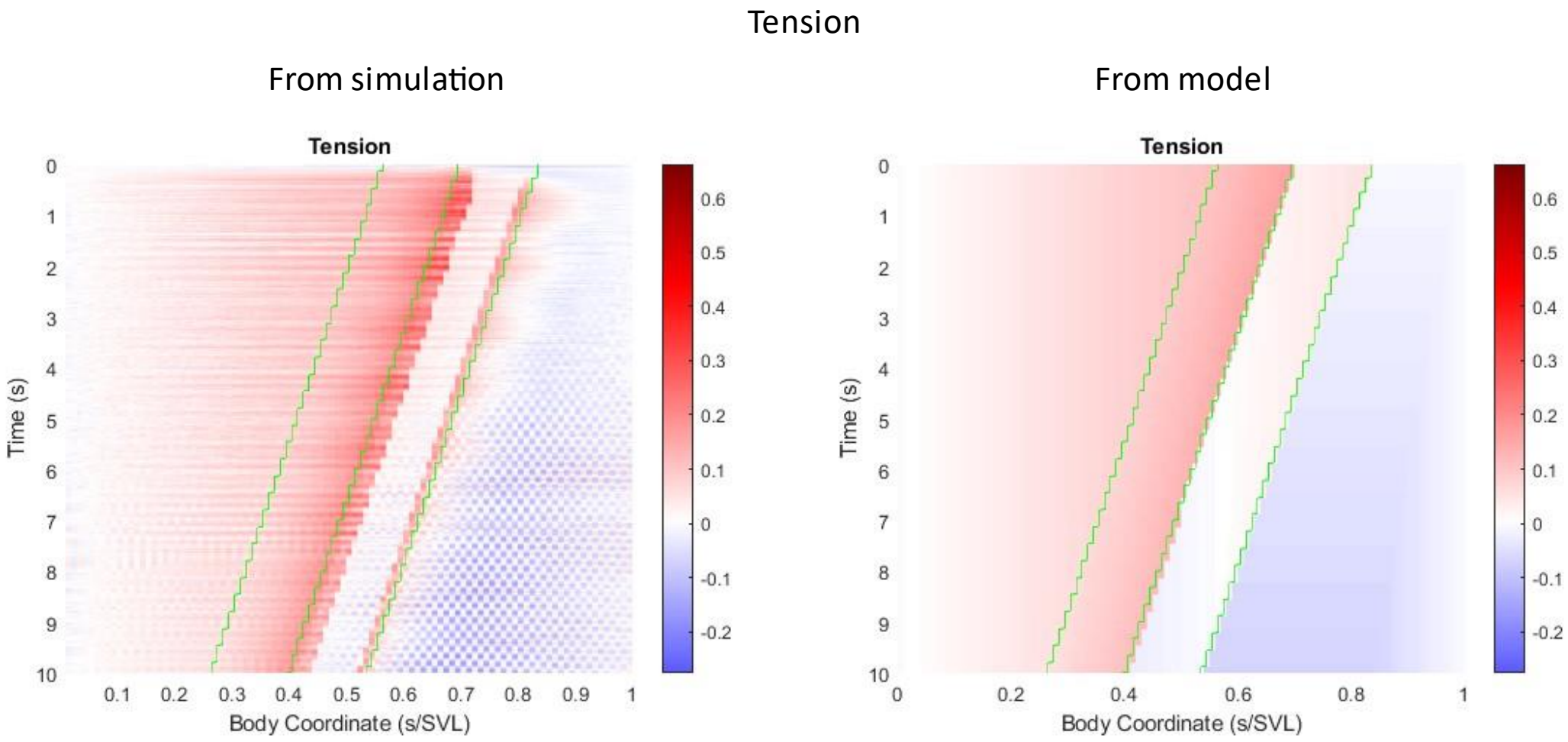

Fig. 9 The tension from simulation results and model analysis. The tension peaks at the inflection point 2 and inflection point 3.

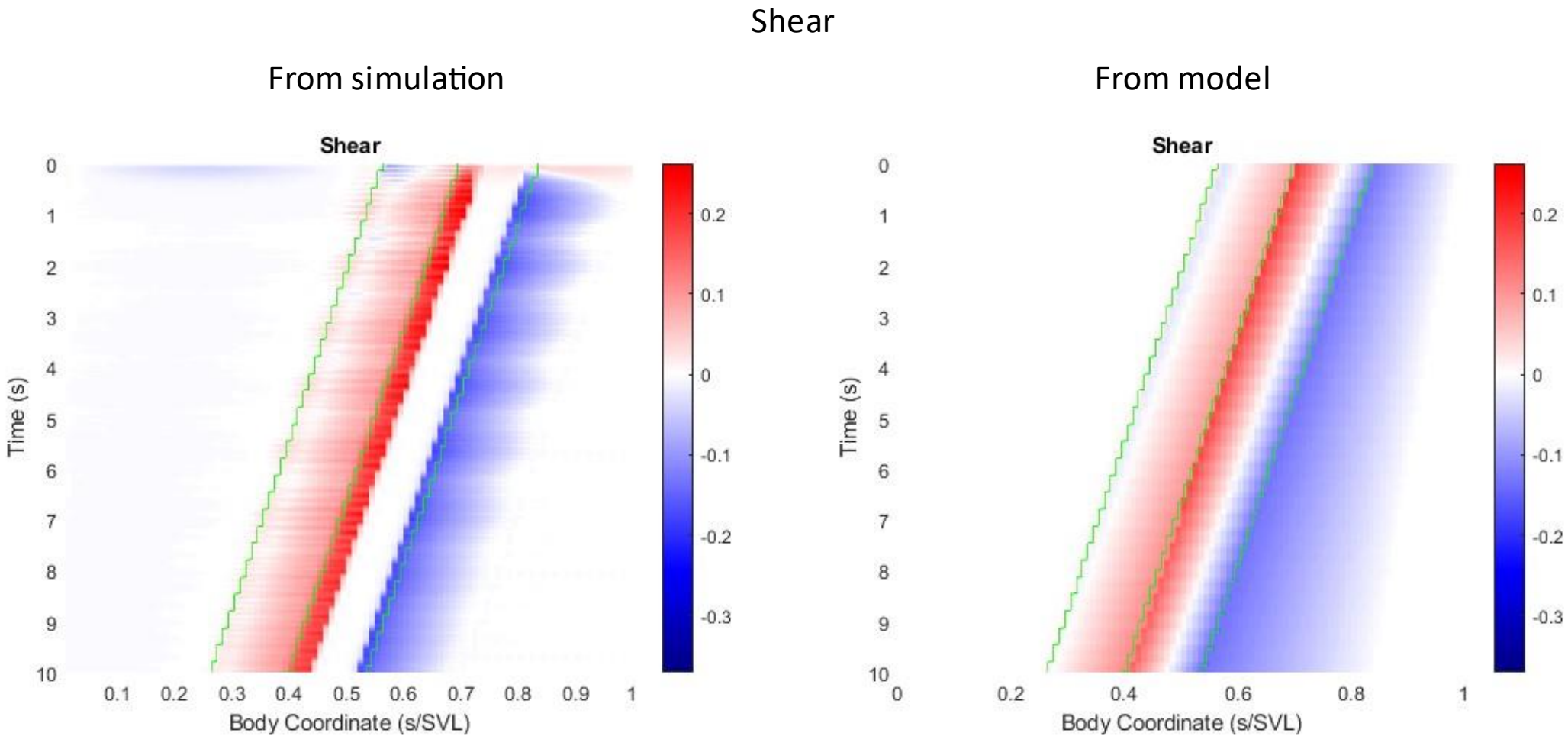

Fig. 10 The shear from simulation results and model analysis. The shear also peaks at the inflection point 2 and inflection point 3.

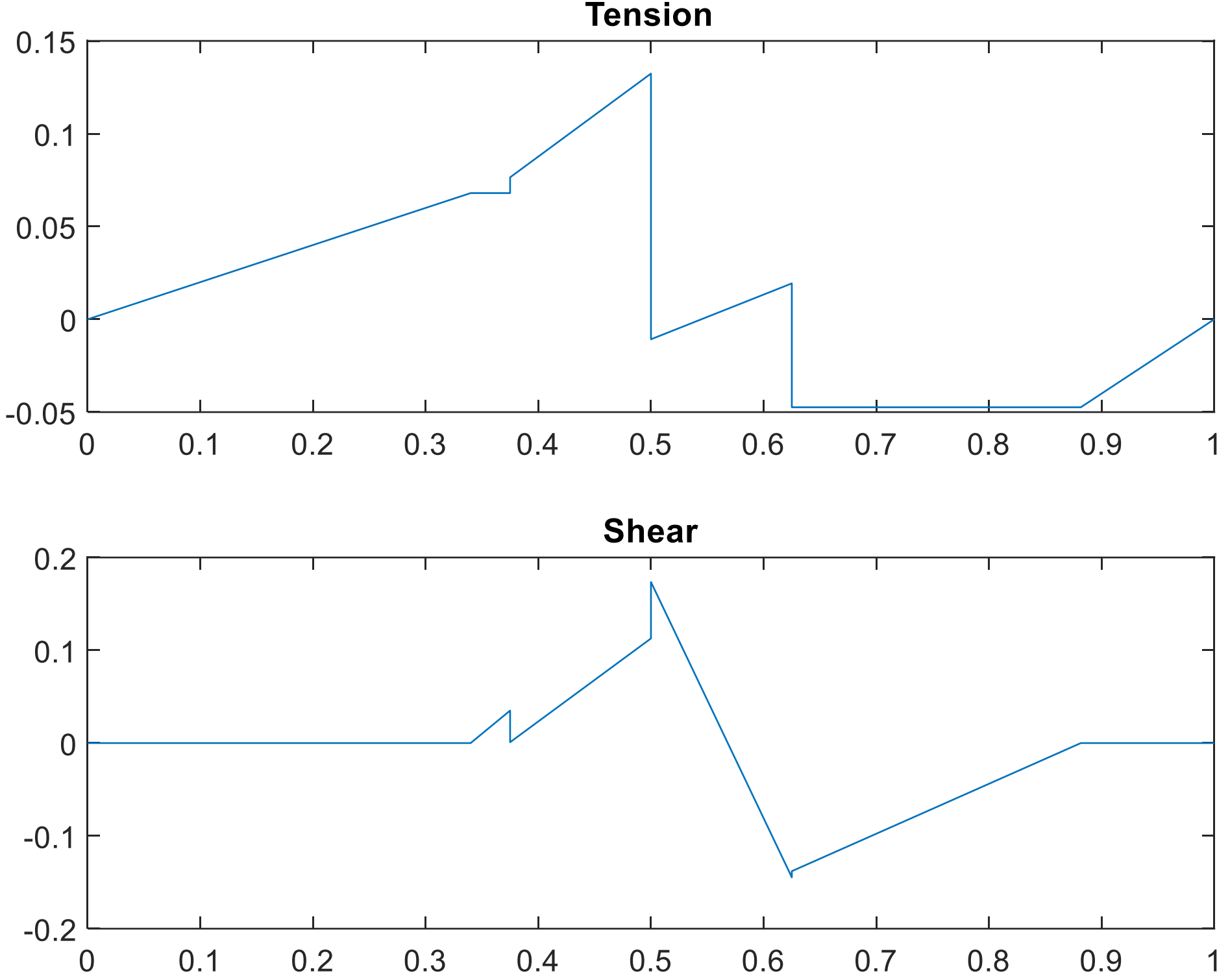

Fig. 11 The schematic of internal force model analysis. It suggests that the internal force peaks around the inflection point 2 and 3.

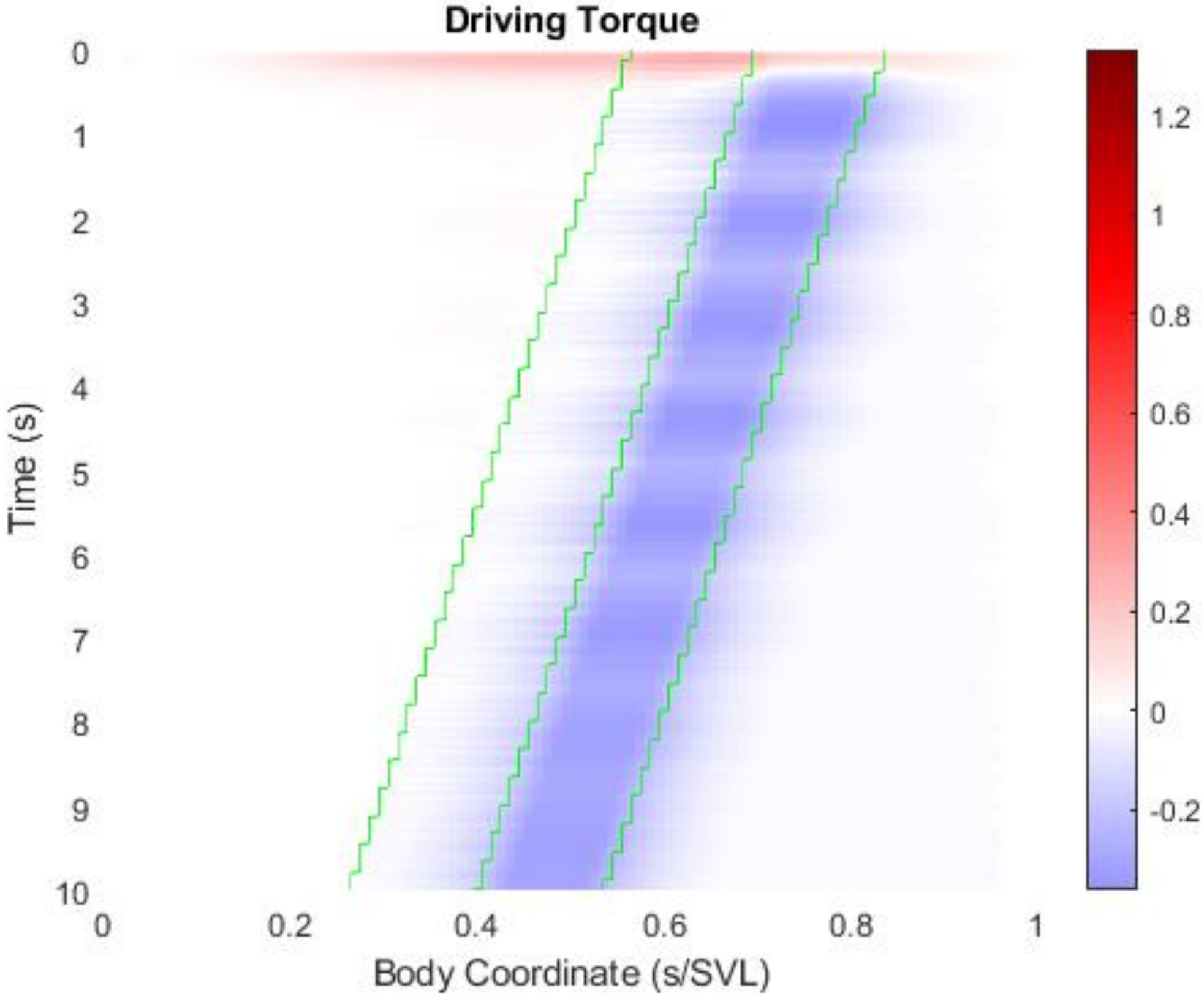

Fig. 12 The driving torque from the simulation results. A posteriorly propagating internal torque profile with a maximum on body segments around the wedge obstacle is displayed in the colormap. Driving torque around the body segments on the flat ground is near zero.

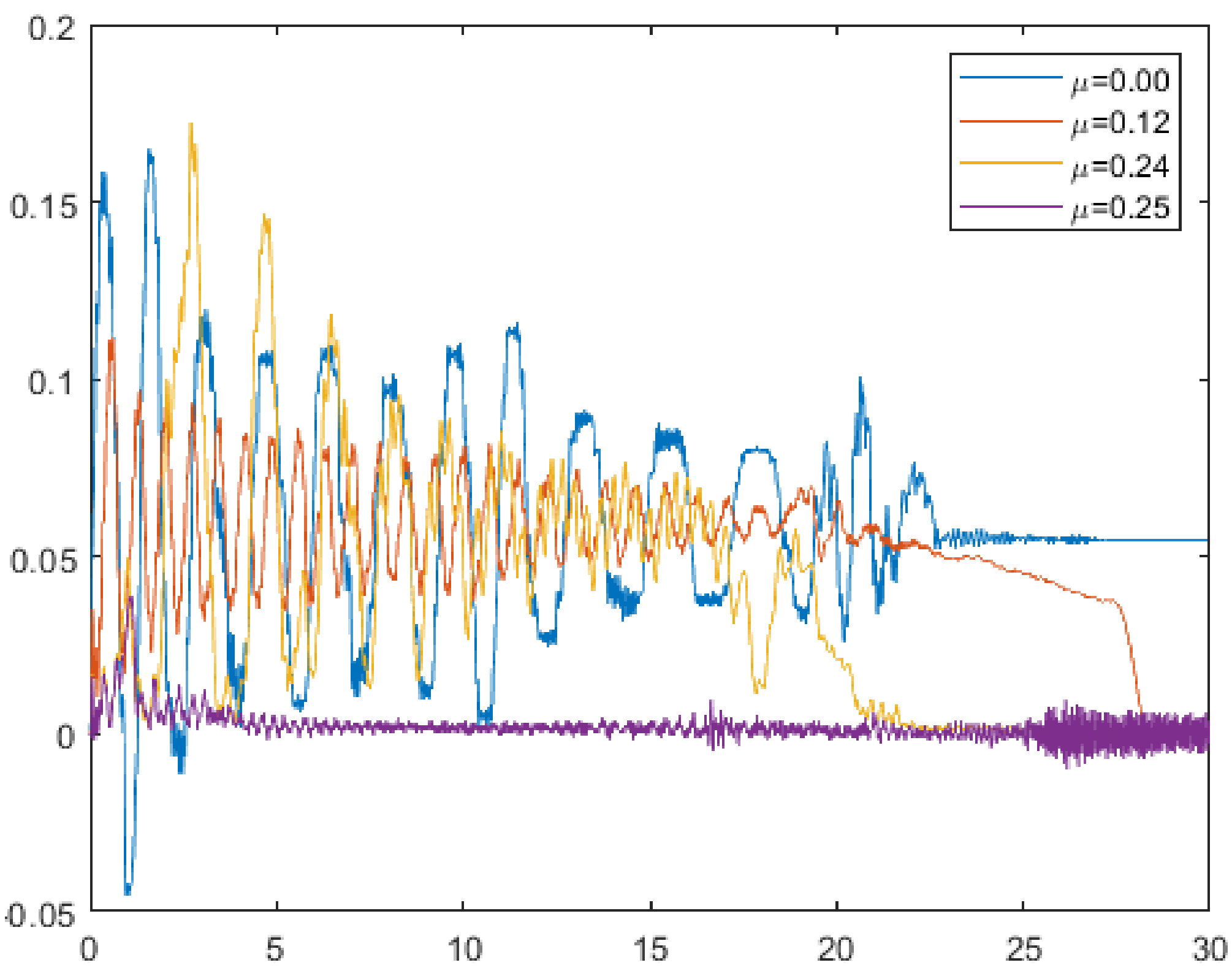

Fig. 13. Speed across different friction coefficients. \mu=0.24 is the largest friction coefficient for a successful traversal, which can be defined as the critical friction coefficient.

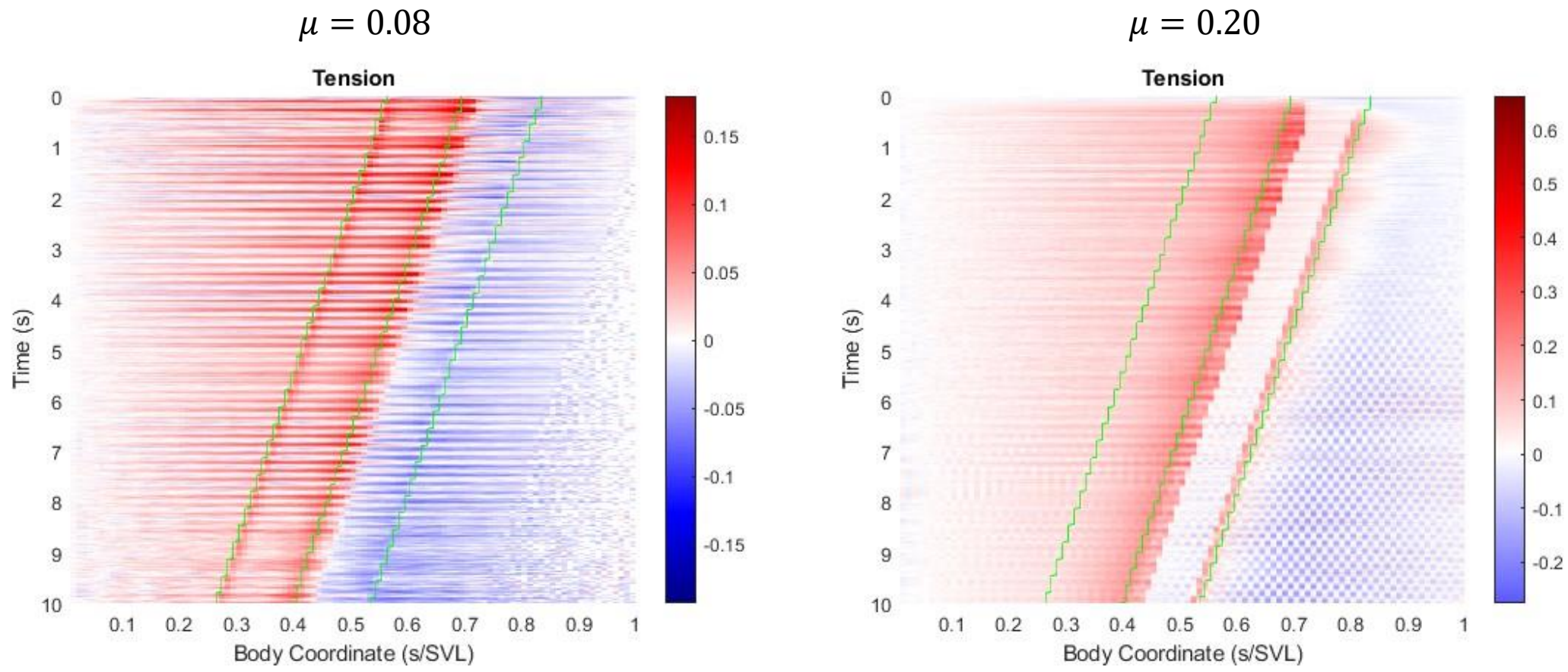

Fig. 14 (a) The comparison of tension force between $\mu = 0.08$ and $\mu = 0.20$. The tension's magnitude is proportional to the friction coefficient.

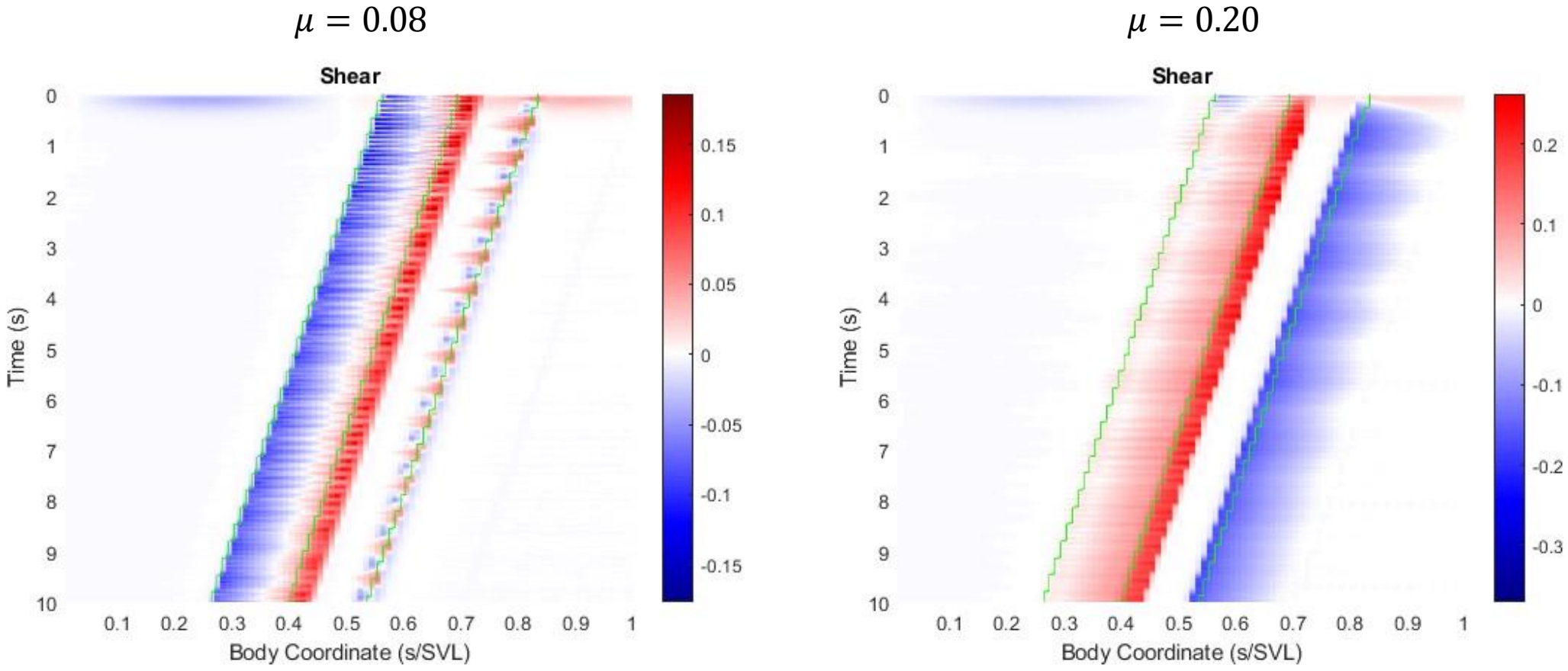

Fig. 14 (b) The comparison of tension force between $\mu = 0.08$ and $\mu = 0.20$. The shear's magnitude is insensitive to the friction coefficient.

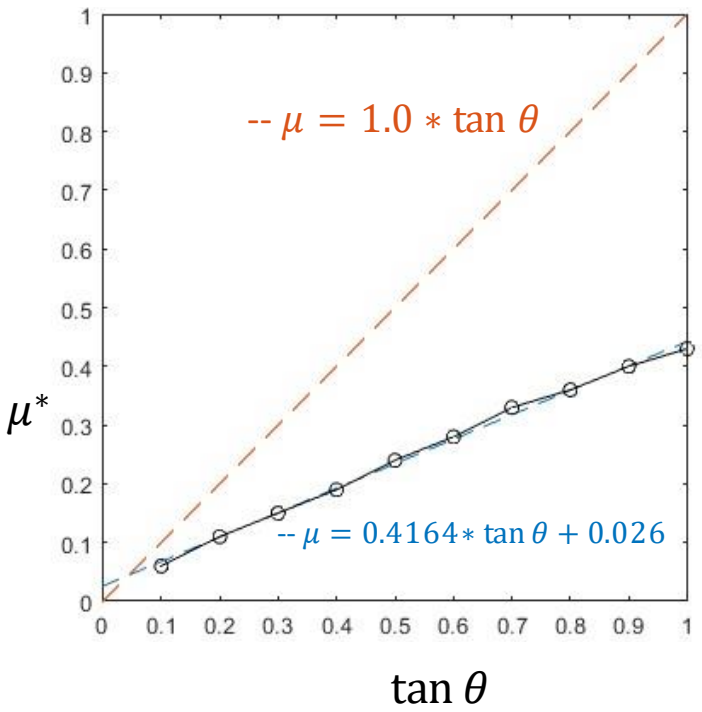

Fig. 15. Critical friction coefficient increases with slope. The critical friction coefficient is well below the slope, mostly around half of it across different slopes.

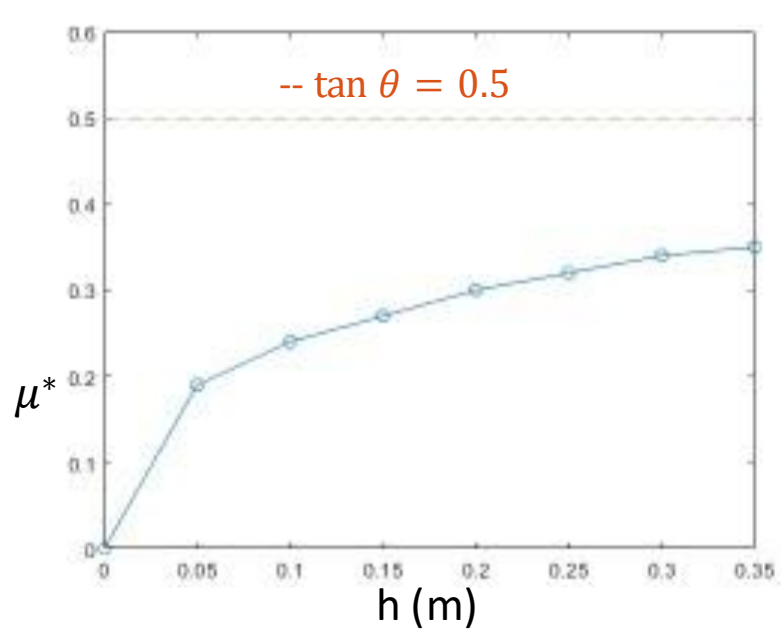

Fig. 16. Critical friction coefficient increases with wedge size. The critical friction coefficient approaches but never exceeds $\tan\theta$.

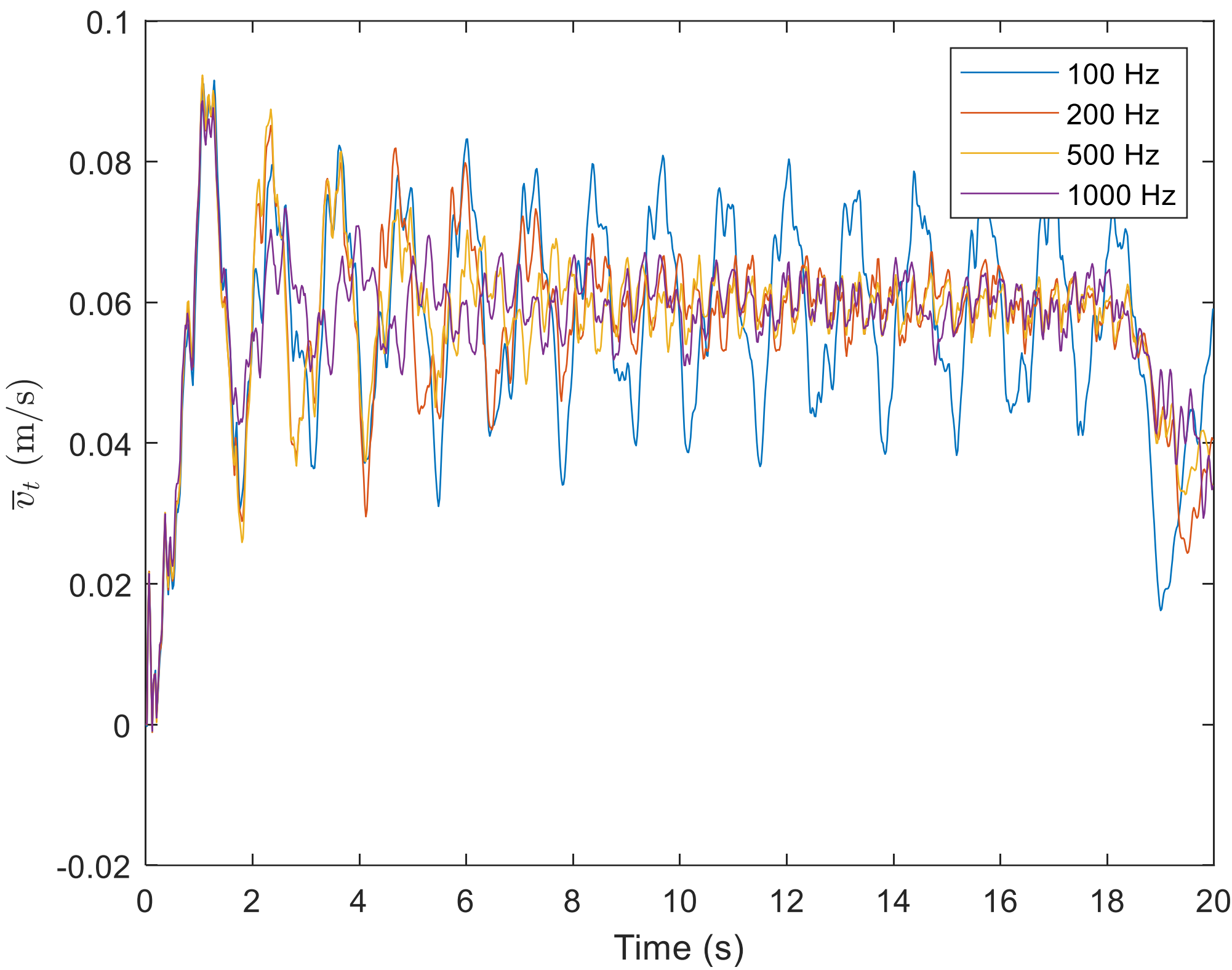

Fig. 17. Comparison of pure propagation gait between different terrain stiffness. Higher stiffness with higher damping has higher speed oscillation frequency and lower amplitude.

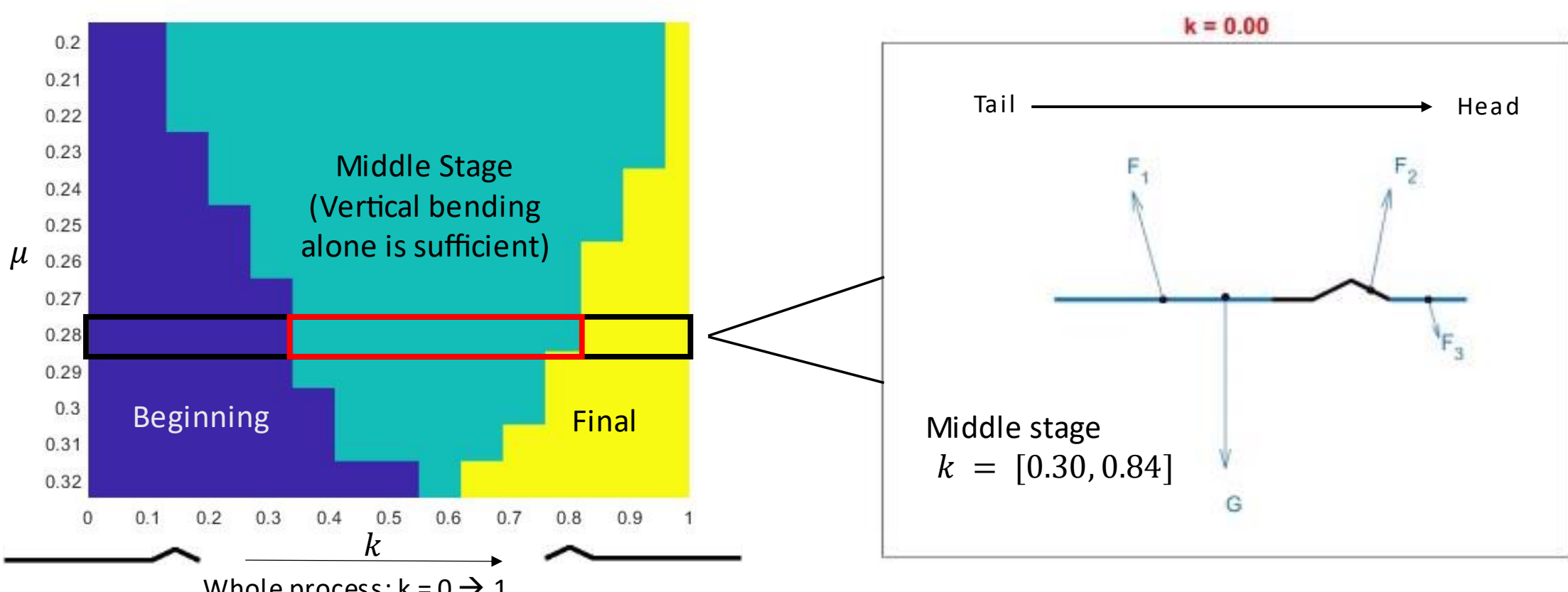

Fig. 18. Different stages of snake traversing a wedge. The algorithm shown in the video can predict the range of the middle stage.

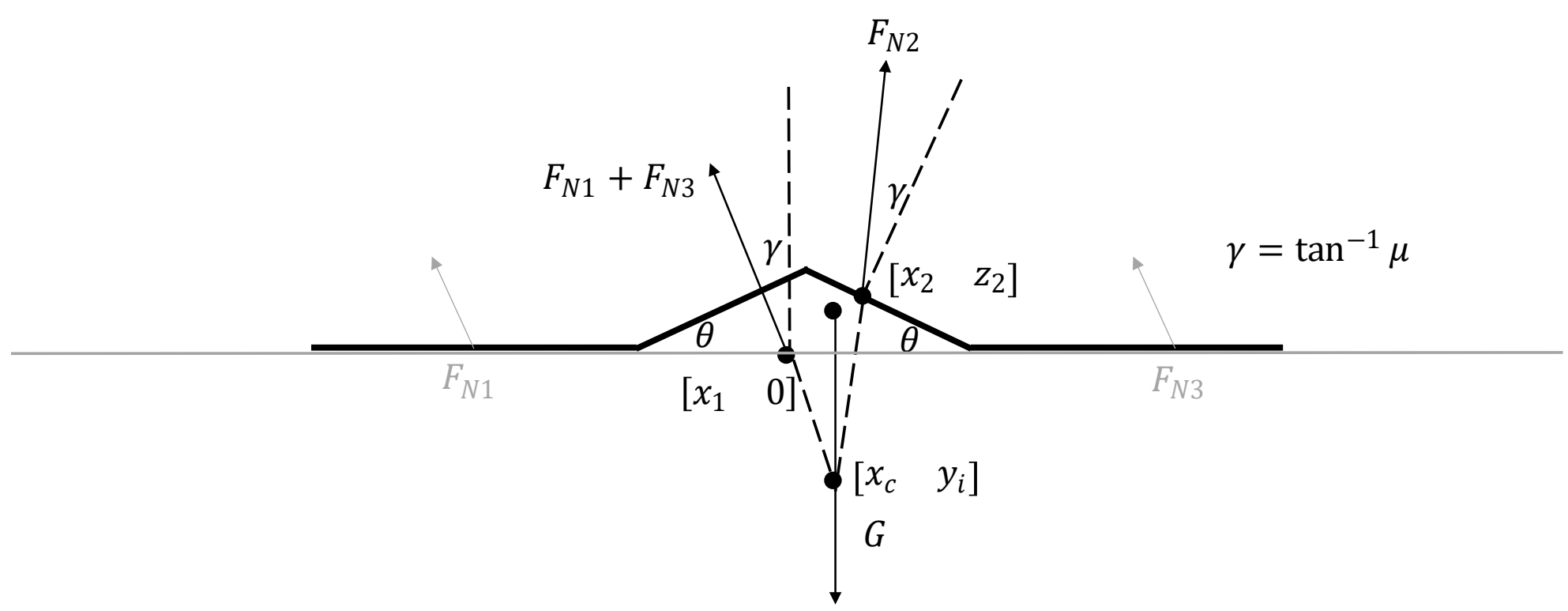

Fig. 19. Torque balance. FN2, the gravitational force and the combination of FN1 and FN3 intersect at [xc, yi].

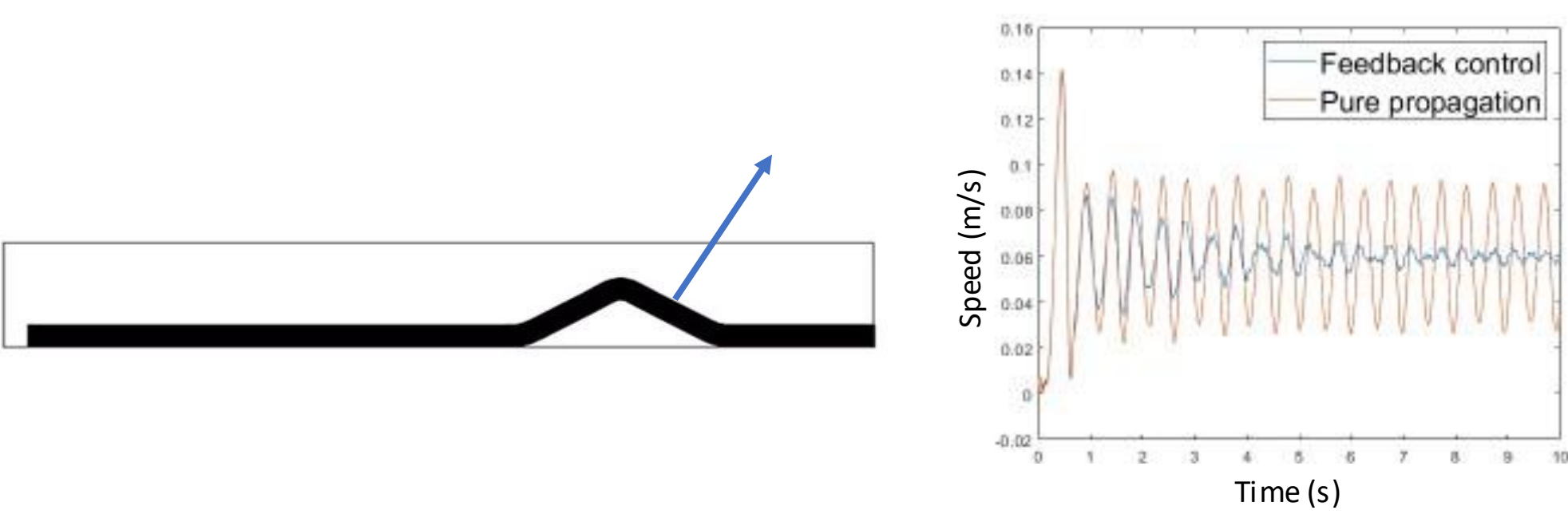

Fig. 20. Force feedback control. The control goal is to track a steady speed. The only source to gain propulsion in this scenario is from the slope. The method is to control the propulsion close to the theoretic value of the steady motion.